%% file: paper.tex
\documentclass[twocolumn,final]{svjour3}          
\smartqed  
\usepackage{graphicx}
\usepackage{multirow}
\usepackage{epsfig}
\usepackage{subfigure}
\usepackage{array}
\usepackage{amsmath}
\usepackage{amssymb}
\usepackage[ruled,vlined,linesnumbered]{algorithm2e}
\usepackage{color}
\usepackage{subfigure}
\usepackage{url}
\usepackage[table,xcdraw]{xcolor}
\usepackage{seqsplit}
\newcommand{\myNum}[1]{(\emph{#1})}
\newcommand{\stitle}[1]{\vspace{1ex}\noindent\textup{\textbf{#1}}}
\newcommand{\nsstitle}[1]{\noindent\textup{\textbf{#1}}}
\newcommand{\logo}{AdHash}

\newcommand{\RDF}[3]{\ensuremath{{\langle}\texttt{#1},}
  \ensuremath{#2,} \ensuremath{\texttt{#3}{\rangle}}}
\newtheorem{myobs}{Observation}
\newtheorem{mydef}{Definition}

\journalname{The VLDB Journal}

\begin{document}

\title{Adaptive Partitioning for Very Large RDF Data}

\author{Razen Harbi \and Ibrahim Abdelaziz \and Panos Kalnis \and Nikos Mamoulis \and Yasser Ebrahim \and Majed Sahli}

\institute{R. Harbi \and I. Abdelaziz \and P. Kalnis \and M. Sahli \at King Abdullah University of Science \& Technology, Thuwal, Saudi Arabia\\
           \email{\{first\}.\{last\}@kaust.edu.sa}         
           \and
           N. Mamoulis \at University of Ioannina, Greece\\
           \email{nikos@cs.uoi.gr}
           \and
           Y. Ebrahim \at Microsoft Corporation, Redmond, WA 98052, United States \\
           \email{yaelsa@microsoft.com}
}

\date{}

\maketitle

\input{sec_abstract}
\input{sec_introduction}
\input{sec_related}

\input{sec_sys_arch}

\input{sec_distributed}

\input{sec_adaptivity}

\input{sec_experimental}
\input{sec_conclusion}
\bibliographystyle{spmpsci}
\bibliography{references}
\input{sec_appendix.tex}
\end{document}

%% file: sec_abstract.tex
\begin{abstract}
State-of-the-art distributed RDF systems partition data across
multiple computer nodes (workers). 
Some systems perform cheap hash partitioning, which may result in
expensive query evaluation, while others apply heuristics aiming at 
minimizing inter-node communication during query evaluation. This
requires an expensive data pre-processing phase, leading to high startup costs
for very large RDF knowledge bases.
Apriori knowledge of the query workload has also been used to 
create partitions, which however are static and 
do not adapt to workload changes; as a result,
inter-node communication 
cannot be consistently avoided 
for queries that are not favored by the initial data partitioning.

In this paper, we propose {\logo}, a distributed RDF system, which
addresses the shortcomings of previous work. 
First, {\logo} 
applies lightweight partitioning on the initial data, that 
distributes triples by hashing on their subjects; this renders its startup overhead low.
At the same time, the locality-aware query optimizer of {\logo} 
takes full advantage of the partitioning to (i)
support the fully parallel processing of join patterns on subjects and
(ii) minimize data communication for general queries by applying hash
distribution of intermediate results instead of broadcasting, wherever possible.
Second, {\logo} monitors the data access patterns and dynamically
redistributes and replicates the instances of the most frequent ones
among workers. As a result, the communication cost for future queries 
is drastically reduced or even eliminated.
To control replication, {\logo} implements 
an eviction policy for the redistributed patterns. 
Our
experiments with synthetic and real data verify that {\logo} (i)
starts faster than all existing systems, (ii) processes thousands of
queries before other systems become online, and (iii) gracefully adapts to the query load, being able to evaluate 
queries on billion-scale RDF data in sub-seconds.

\end{abstract}

%% file: sec_introduction.tex
\section{Introduction}
\label{sec:introduction}

\begin{figure}
  \centering
  \includegraphics[width=.9\columnwidth]{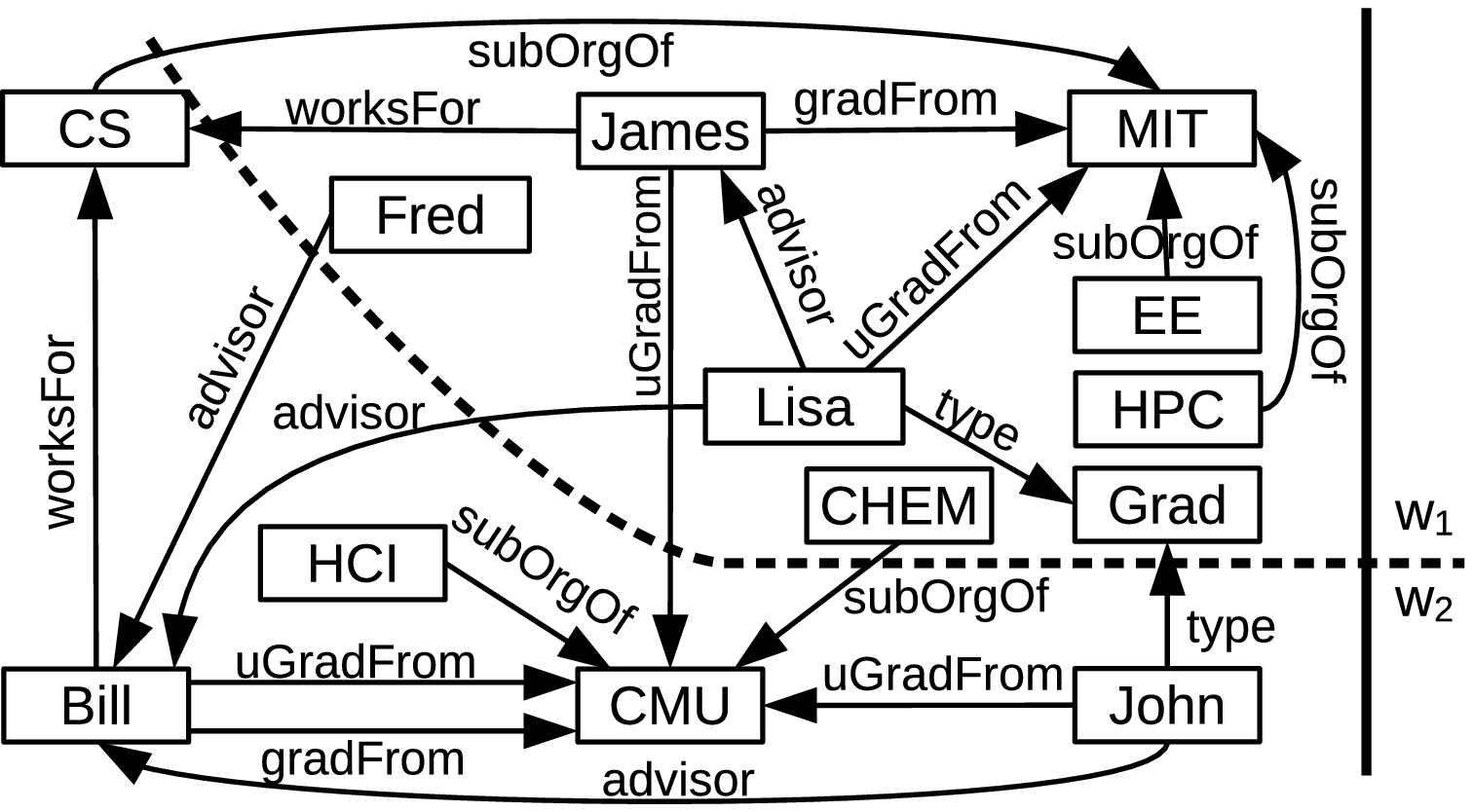}
  \caption{Example RDF graph. An edge and its associated vertices correspond to an RDF triple; e.g., \RDF{Bill}{worksFor}{CS}.}
  \label{fig:rdf_graph}
\end{figure}

The RDF data model does not require a predefined schema and 
represents information from diverse sources in a versatile manner.
Therefore, social networks, search engines, shopping sites and scientific databases are adopting RDF for publishing Web content. Many large public knowledge bases, such as Bio2RDF\footnote{http://www.bio2rdf.org/} and YAGO\footnote{http://yago-knowledge.org/} have billions of facts in RDF format. 
RDF datasets consist of  triples of the form \RDF{subject}{predicate}{object}, where $predicate$ represents a relationship between two entities: a \verb|subject| and an \verb|object|. 
An RDF dataset can be regarded as a long relational table with three columns. An RDF dataset  can also be viewed as a directed labeled graph, where vertices and edge labels correspond to entities and predicates, respectively. Figure~\ref{fig:rdf_graph} shows an example RDF graph of students and professors in an \mbox{academic network.}

\begin{figure}
  \centering
  \includegraphics[width=.8\columnwidth]{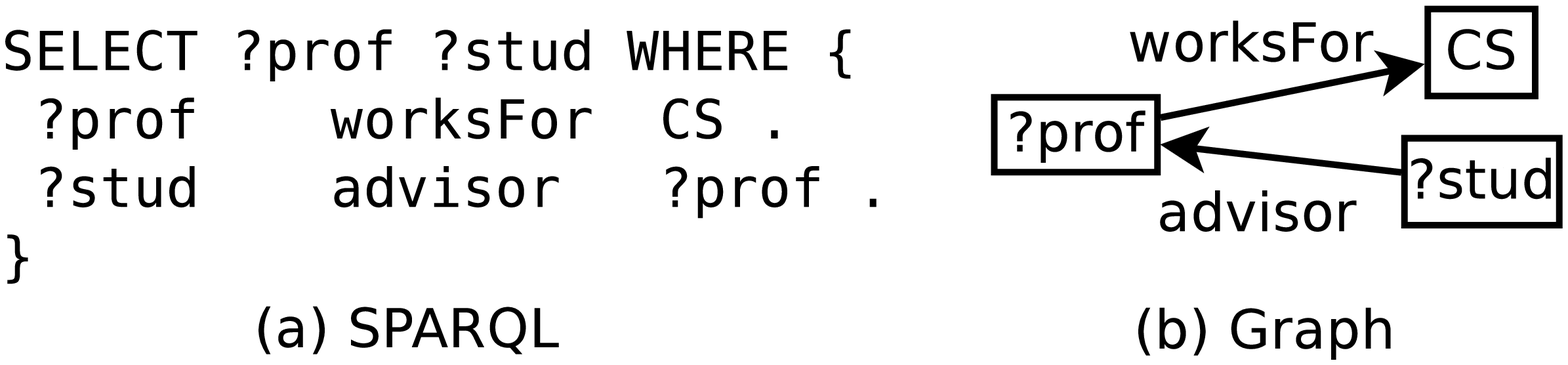}
  \caption{A query that finds CS professors with their advisees.}
  \label{fig:queryProf} 
\end{figure}

SPARQL\footnote{http://www.w3.org/TR/rdf-sparql-query/} is the standard query language for RDF. 
Each query is a set of RDF triple patterns; 
some of the nodes in a pattern are variables 
which may appear in multiple patterns.
For example, the query in Figure \ref{fig:queryProf}(a) returns all professors who work for CS with their advisees. The query corresponds to the graph pattern in Figure~\ref{fig:queryProf}(b). The answer is the set of ordered bindings of $(?p, ?s)$ that render the query graph isomorphic to subgraphs in the data. Assuming the data is stored in a table $D(s,p,o)$, the query can be answered by first decomposing it into two subqueries, each corresponding to a triple pattern: 
$q_1 \equiv \sigma_{p=worksFor \wedge o=CS}(D)$ and
$q_2 \equiv \sigma_{p=advisor}(D)$.
The subqueries can be answered independently by scanning table $D$;
then, we can join their intermediate results on the subject and object attribute: 
$q_1\bowtie_{q_1.s = q_2.o} q_2$. By applying the query on the data of Figure \ref{fig:rdf_graph}, 
we get \texttt{(?prof, ?stud)} $\in \{$\texttt{(James, Lisa),(Bill, John), (Bill, Fred),(Bill, Lisa)}$\}$.

As the volume of RDF data continues soaring, managing, indexing and querying
RDF data collections becomes challenging. Early research efforts focused on building efficient centralized RDF systems; like RDF-3X \cite{Neumann2010}, HexaStore \cite{Weiss2008}, TripleBit \cite{triplebit} and gStore \cite{gstore}.
However, centralized data management and search does not scale well for complex queries on web-scale RDF data.
As a result, distributed RDF management systems were introduced to improve performance. Such systems scale-out by partitioning RDF data among many computer nodes (workers) and evaluating queries in a distributed fashion. A SPARQL query is decomposed into multiple subqueries that are evaluated by each node independently. Since data is distributed, the nodes may need to exchange intermediate results during query evaluation. Consequently, queries with large intermediate results incur high communication cost, which is detrimental to the query performance \cite{triad,Huang11:scalable}.

Distributed RDF systems aim at minimizing the number of decomposed subqueries by partitioning the data among workers. The goal is that each node has all the data it needs to evaluate the entire query and there is no need for exchanging intermediate results. In such a {\em parallel} query evaluation, each node contributes a partial result of the query; the final query result is the union of all partial results. To achieve this, some triples may need to be replicated in multiple partitions. For example, in Figure \ref{fig:rdf_graph}, assume the data graph is divided by the dotted line into two partitions and assume that triples follow their subject placement. To answer the query in Figure \ref{fig:queryProf}, nodes have to exchange intermediate results because triples \RDF{Lisa}{advisor}{Bill} and \RDF{Fred}{advisor}{Bill} cross the partition boundary. Replicating these triples in both partitions allows each node to answer the query without communication. Still, even sophisticated partitioning and replication cannot guarantee that arbitrarily complex SPARQL queries can be processed in parallel; thus, expensive {\em distributed} query evaluation, with intermediate results exchanged between nodes cannot always be avoided. 

\stitle{Challenges.} Existing distributed RDF systems are facing two limitations. \myNum{i} \emph{Partitioning cost:} graph partitioning is an NP-complete problem \cite{metis}; thus, existing systems perform heuristic partitioning. 
In systems like \cite{Harth07yars2,h2rdf2,Rohloff10:shard,trinity.rdf} 
that use simple hash partitioning heuristics,
queries have low chances to be evaluated in parallel without any communication between nodes.
On the other hand, systems that use sophisticated partitioning heuristics \cite{triad,Huang11:scalable,shape,path} suffer from high preprocessing cost and sometimes high replication. More importantly, they pay the cost of partitioning the entire data regardless of the anticipated workloads. However, as shown in a recent study \cite{sampled}, only a small fraction of the whole graph is actually accessed by typical real query workloads. For example, a real workload consisting of more than 1,600 queries executed on DBpedia (459M triples) touches only 0.003\% of the whole data. Therefore, we argue that distributed RDF systems should leverage query workloads in data partitioning. 
\myNum{ii} \emph{Adaptivity:} WARP \cite{warp} and Partout \cite{partout} do consider the workload during data partitioning and achieve 
a significant reduction in the replication ratio, while showing better query performance compared to systems that partition the data blindly. Nonetheless, both these systems assume a representative (i.e., {\em static}) query workload and do not adapt to changes. 
However, because of workloads diversity and dynamism, Alu\c{c} et al. \cite{wlm} showed that systems need to continuously adapt to workloads in order to consistently provide good performance; relying on a static workload results in performance degradation for queries that are not represented by it. 

In this paper, we propose \textbf{Ad}aptive \textbf{Hash}ing ({\logo}), a distributed in-memory RDF engine. {\logo} alleviates the aforementioned limitations of existing systems based on the following key principles. \\
\textbf{Lightweight Initial Partitioning:} {\logo} uses an initial hash partitioning,
that distributes triples by hashing on their subjects. 
This partitioning has low cost and does not incur any replication.
Thus, the preprocessing time is low, partially addressing the \mbox{first challenge.} \\
\textbf{Hash-based Locality Awareness:} {\logo} 
achieves competitive performance by  maximizing the number of joins that can be executed in parallel without data communication by exploiting hash-based locality; 
the join patterns on subjects included in a query can be processed in parallel.
In addition, intermediate results can potentially be hash-distributed to single workers instead of being broadcasted everywhere.
The locality-aware query optimizer of {\logo}  
considers these properties to compute an evaluation plan that minimizes intermediate results shipped between workers.\\
\textbf{Adapting by Incremental Redistribution:} {\logo} monitors the executed workload and incrementally updates a hierarchical heat-map of accessed data patterns. Hot patterns are redistributed and potentially replicated in the system in a way that future queries that include them are executed in parallel by all workers without data communication. To control replication, {\logo} operates within a budget and employs an eviction policy for the redistributed patterns. This way, {\logo} overcomes the limitations of static partitioning schemes and adapts dynamically to changing workloads. 

In summary, our contributions are:
\begin{itemize}
\item We introduce {\logo}, a distributed SPARQL engine that does not require expensive preprocessing. By using lightweight hash partitioning, avoiding the upfront cost, and adopting a pay-as-you-go approach, 
{\logo} executes tens of thousands of queries on large graphs within the time it takes other systems to conduct their initial partitioning.

\item We propose a locality-aware query planner and a cost-based optimizer for {\logo} to efficiently execute queries that require data communication.

\item We present a novel approach for monitoring and indexing workloads in the form of hierarchical heat maps. Queries are transformed and indexed using these maps to facilitate the adaptivity of {\logo}.
We introduce an Incremental ReDistribution (IRD) technique. Guided by the workload, IRD incrementally redistributes portions of the data that are accessed by hot patterns. Based on IRD, 
{\logo} processes future queries without data communication.

\item We evaluate {\logo} using synthetic and real data and compare with state-of-the-art systems. {\logo} partitions billion-scale RDF data
and starts answering queries in less than 14 minutes, while other systems need hours or days. {\logo} executes large workloads orders of magnitude faster than existing approaches. To the best of our knowledge, {\logo} is the only system capable of providing sub-second execution times for queries with complex structures on billion scale RDF data.

\end{itemize}

The rest of the paper is organized as follows. Section \ref{sec:relatedWork} reviews existing distributed RDF systems and the techniques used by them for scalable SPARQL query evaluation. Section \ref{sec:sys_arch} presents the architecture of {\logo} and provides an overview of the system's components. Section \ref{sec:distributed_query} discuses our locality-aware query planning and distributed query evaluation, whereas Section \ref{sec:adaptivity} explains the adaptivity feature of {\logo}. Section \ref{sec:experimental} contains the experimental results and Section \ref{sec:conclusion} concludes the paper.

%% file: sec_related.tex
\section{Related Work}
\label{sec:relatedWork}
In this section, we review recent distributed RDF systems, which are related to AdHash. 
Table \ref{tab:related_summary}, summarizes the main characteristics of these systems.
\begin{table*}[t]\scriptsize
\centering
\caption{Summary of state-of-the-art distributed RDF systems}
\begin{tabular}{llcccc}
\hline
\multicolumn{1}{c}{System}          & \multicolumn{1}{c}{\begin{tabular}[c]{@{}c@{}}Partitioning\\ Strategy\end{tabular}}         & \multicolumn{1}{c}{\begin{tabular}[c]{@{}c@{}}Partitioning\\ Cost\end{tabular}} & \multicolumn{1}{c}{Replication} & \multicolumn{1}{c}{\begin{tabular}[c]{@{}c@{}}Workload\\ Awareness\end{tabular}} & \multicolumn{1}{c}{Adaptive} \\\hline
TriAD \cite{triad}                  & Graph-based (METIS) \& Horizontal triple Sharding                                           & High                                                                            & Yes                             & No                                                                               & No                           \\
H-RDF-3X \cite{Huang11:scalable}    & Graph-based (METIS)                                                                         & High                                                                            & Yes                             & No                                                                               & No                           \\
Partout \cite{partout}              & Workload-based horizontal fragmentation                                                     & High                                                                            & No                              & Yes                                                                              & No                           \\
SHAPE \cite{shape}            	    & Semantic Hash                                                                               & High                                                                        & Yes                             & No                                                                               & No                           \\
Wu et al. \cite{path}               & End-to-end path partitioning                                                                & Moderate                                                                        & Yes                             & No                                                                               & No                           \\
Trinity.RDF \cite{trinity.rdf}      & Hash                                                                                        & Low                                                                             & Yes                             & No                                                                               & No                           \\
H2RDF+ \cite{h2rdf2}                & H-Base partitioner (range)                                                                  & Low                                                                             & No                              & No                                                                               & No                           \\
SHARD  \cite{Rohloff10:shard}       & Hash                                                                                        & Low                                                                             & No                              & No                                                                               & No                           \\
AdHash                              & Hash                                                                                        & Low                                                                             & Yes                             & Yes                                                                              & Yes                          \\ \hline
\end{tabular}
\label{tab:related_summary}
\end{table*}

\stitle{Lightweight Data Partitioning: }
Several systems are based on the MapReduce framework \cite{mapred} and use the  Hadoop Distributed File System (HDFS) to store RDF data. 
HDFS uses 
horizontal random data partitioning.
SHARD \cite{Rohloff10:shard} stores the whole RDF data into one HDFS  file. 
HadoopRDF \cite{hadooprdf} also uses HDFS but splits the data into multiple smaller files. 
SHARD and HadoopRDF solve SPARQL queries using a set of MapReduce iterations.

Trinity.RDF \cite{trinity.rdf} is a distributed in-memory RDF engine that can handle web scale RDF data. 
It represents RDF data in its native graph form (i.e., using adjacency lists) and uses a key-value store as the back-end storage. 
The RDF graph is partitioned using vertex id as hash key. This is equivalent to partitioning the data twice; first using subjects 
as hash keys and second using objects. 
Trinity.RDF uses {\em graph exploration} for SPARQL query evaluation and relies heavily on its underlying high-end \mbox{InfiniBand} interconnect. 
In every iteration, a single subquery is explored starting from valid bindings by all workers. 
This way, generation of redundant intermediate results is avoided. However, because exploration only involves two vertices (source and target), Trinity.RDF cannot 
prune invalid intermediate results without carrying all their historical bindings. Hence, workers need to ship candidate results to the master 
to finalize the results, which is a potential bottleneck of the system.

Rya \cite{rya} and H2RDF+ \cite{h2rdf2} use key-value stores for RDF data storage which range-partition the data based on keys such that 
the keys in each partition are sorted. 
When solving a SPARQL query, Rya executes the first subquery using range scan on the appropriate index; it then utilizes index lookups for the next subqueries. 
H2RDF+ executes simple queries in a centralized fashion, whereas complex queries are solved using a set of MapReduce iterations. 

All the above systems use lightweight partitioning schemes, which are computationally inexpensive; however, queries with long paths and complex structures incur high communication costs. In addition, systems that use MapReduce for join evaluation suffer from its high overhead \cite{triad,path}.
On the contrary, although our {\logo} system also uses lightweight hash partitioning, it avoids excessive data shuffling by exploiting  hash-based data locality. Furthermore, it adapts incrementally to the workload to further minimize communication.

\stitle{Sophisticated Partitioning Schemes and Replication: }
Several systems employ general graph partitioning techniques to partition RDF data, in order to improve data locality. 
EAGRE \cite{eagre} focuses on minimizing the I/O cost. 
The RDF graph is transformed into a compressed entity graph that is partitioned using a MinCut algorithm, such as METIS \cite{metis}. 
H-RDF-3X \cite{Huang11:scalable} uses METIS to partition the RDF graph among workers. 
It also enforces the so-called $k$-hop guarantee so any query with radius $k$ or less can be executed without communication. Queries with radius larger than $k$ are executed using expensive MapReduce joins. Replication increases exponentially with $k$; therefore, $k$ must be kept small (e.g., $k\leq2$ in \cite{Huang11:scalable}). Both EAGRE and H-RDF-3X suffer from the significant overhead of MapRe\-duce-based joins for queries that cannot be evaluated locally. For such queries, sub-second query evaluation is not possible \cite{triad}, even with state-of-the-art MapReduce implementations, like Hadoop++ \cite{hadooppp} and Spark \cite{spark}.

TriAD \cite{triad} uses METIS for data partitioning. Edges which cross partitions replicated resulting in a $1{-}hop$ guarantee. 
A summary graph is defined, which includes a vertex for each partition. Vertices in this graph are connected by the cross-partition edges. 
A query in TriAD is evaluated against the summary graph first, in order to prune partitions that do not contribute to query results. 
Then, the query is evaluated on the RDF data residing in the partitions retrieved from the summary graph. 
Multiple join operators are executed concurrently by all workers, which communicate via an asynchronous message passing protocol. 
Sophisticated partitioning techniques, like MinCut, reduce the communication cost significantly. 
However, such techniques are prohibitively expensive and do not scale for large graphs, as shown in \cite{shape}.
Furthermore, MinCut does not yield good partitioning for dense graphs. 
Thus, TriAD does not benefit from the summary graph pruning technique in dense RDF graphs because of the high edge-cut.
To alleviate METIS overhead, an efficient approach for partitioning large graphs was introduced \cite{billion}.  
Nonetheless, 
there will always be SPARQL queries with poor locality that cross partition boundaries and result in poor performance.

SHAPE \cite{shape} proposed a semantic hash portioning approach for RDF data. SHAPE starts by simple hash partitioning and employs the same $k$-hop strategy as H-RDF-3X \cite{Huang11:scalable}. It also relies on URI hierarchy, for grouping vertices to increase data locality. Similar to H-RDF-3X, SHAPE suffers from the high overhead of MapReduce-based joins. 
Furthermore, URI-based grouping results in skewed partitioning if a large percentage of vertices share prefixes. This behavior is noticed in both real as well as synthetic datasets (See Section \ref{sec:experimental}).

Recently, Wu et al. \cite{path} proposed an end-to-end path partitioning scheme,
which considers all possible directed paths in the RDF graph. These paths are merged in a bottom-up fashion, beginning with the paths starting vertices. While this approach works well for star, chain and directed cyclic queries; other types of queries result in significant communication. For example, queries with object-object joins or queries that do not associate each query vertex with the type predicate would require inter-worker communication. Note that our adaptivity technique (Section \ref{sec:adaptivity}) is orthogonal to and can be combined with  
end-to-end path partitioning as well as other partitioning heuristics
to efficiently evaluate queries that are not favored by the partitioning.

\stitle{Workload-Aware Data Partitioning: }
Most of the aforementioned partitioning techniques focus on minimizing communication without considering the workload. 
A recent study \cite{sampled} shows that real query workloads touch a small fraction of the data. 
Therefore, utilizing the query workload helps to reduce 
communication costs for queries that cannot be evaluated in parallel,  
based on the  partitioning scheme used.
Partout \cite{partout} is a distributed engine, which relies on a 
given workload to divide the data between nodes.
It first extracts a representative triple patterns from the query load. Then uses these patterns to partition the data into fragments and collocates fragments that are accessed together by queries on the same worker.
Similarly, WARP \cite{warp} uses a representative query workload to replicate frequently accessed data. 
However, if the workload changes or the user query is not in the representative workload, Partout and WARP incur high communication costs. They can only adapt to changes in the workload, by applying expensive re-partitioning of the entire data.
On the contrary, our {\logo} system adapts incrementally by replicating only the data accessed by the workload which is small, as we discussed.

\nsstitle{SPARQL on Vertex-centric.} 
Sedge \cite{yang12} solves the problem of dynamic graph partitioning and demonstrates its partitioning effectiveness using SPARQL queries over RDF. 
The entire graph is replicated several times and each replica is partitioned differently.
Every SPARQL query is translated manually into a Pregel \cite{pregel10} program and is executed against the replica that minimizes communication. 
Still, this approach incurs excessive replication, as it duplicates the entire data several times. Moreover, its lack of support for ad-hoc queries makes it counter-productive; a user needs to manually write an optimized query evaluation program in Pregel.

\stitle{Materialized views: }
Several works attempt to speed up the execution of SPARQL queries by materializing a set of views \cite{Chong,Goasdoue} or a set of path expressions \cite{Dritsou}. 
The selection of views is based on a representative workload. Our approach does not generate local materialized views. Instead, we redistribute the data 
accessed by hot patterns in a way that preserves data locality and allows queries to be executed with minimal communication.  

\stitle{Relational Model: }
There also exist relevant systems that focus on data models other than RDF. Schism \cite{schism10} deals with data placement for distributed OLTP RDBMS. 
Using a sample workload, Schism minimizes the number of distributed transactions by populating a graph of co-accessed tuples. 
Tuples accessed in the same transaction are put in the same server. 
This is not appropriate for SPARQL because some queries access large parts of the data that would overwhelm a single machine. 
Instead, {\logo} exploits parallelism by executing such a query across all machines in parallel without communication. 
H-Store \cite{stonebraker07} is an in-memory distributed OLTP RDBMS that uses a data partitioning technique similar to ours. 
Nevertheless, H-Store assumes that the schema and the query workload are given in advance and assumes no ad-hoc queries. 
Although, these could be valid assumptions for OLTP databases, they are not for RDF data stores.

\stitle{Eventual indexing: }
Idreos et al. \cite{IdreosKM07} introduced the concept of reducing the data-to-query time for relational data. 
They avoid building indices during data loading; instead, they reorder tuples incrementally during query processing. 
In {\logo}, we extend eventual indexing to dynamic and adaptive graph partitioning. 
In our problem, graph partitioning is very expensive; hence, the potential benefits of minimizing the data-to-query time are substantial.

%% file: sec_sys_arch.tex
\vspace{-6pt}
\section{System Architecture}
\label{sec:sys_arch}

\begin{figure}
  \centering
  \includegraphics[width=1.0\columnwidth]{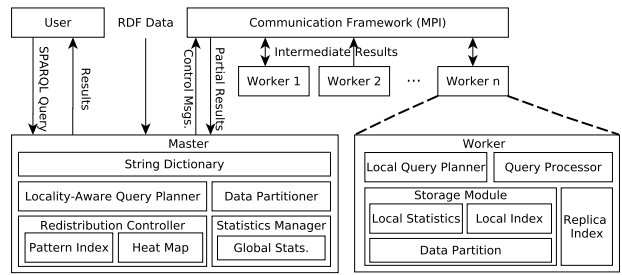}
  \caption{System architecture of {\logo}}
  \label{fig:sys_arch}
\end{figure}

{\logo} employs the typical master-slave paradigm and is deployed on a shared-nothing cluster of machines (see Figure \ref{fig:sys_arch}). The master and workers communicate through message passing.
The same architecture is used by other systems, e.g., Trinity.RDF \cite{trinity.rdf} and TriAD \cite{triad}.

\vspace{-11pt}
\subsection{Master}

The master begins by partitioning the data among workers and collecting global statistics. 
Then, it receives queries from users, generates execution plans, coordinates workers, collects results, and returns final results.

\nsstitle{String Dictionary.} RDF data contains long strings in the form of URIs and literals. To avoid the storage, processing, and communication overheads, we encode RDF strings into numerical IDs and build a bi-directional dictionary. This approach is used by state-of-the-art systems \cite{triad,Neumann2010,h2rdf2,trinity.rdf}. 

\nsstitle{Data Partitioner.} 
A recent study \cite{empirical} showed that joins on the subject column account for 60\% of the joins in a real workload of SPARQL queries. Therefore, {\logo} uses lightweight hash-based triple sharding on subject values. Given $W$ workers, a triple $t$ is assigned to worker $w_i$, where $i$
is the result of a hash function applied on $t.subject$.%
\footnote{For simplicity, we use: $i = t.subject \mod W$.} This way all triples that share the same subject will be assigned to the same worker. Consequently, any star query joining on subjects can be evaluated without communication among workers.
We do not hash on objects because they can be literals and common types. Hashing on objects would assign all triples of the same type to one worker, resulting in load imbalance and limited parallelism \cite{Huang11:scalable}. To validate our argument, we use the synthetic LUBM-4000\footnote{http://swat.cse.lehigh.edu/projects/lubm/} and real YAGO2\footnote{http://yago-knowledge.org/} datasets, which have around 500M and 300M triples, respectively. Both datasets are partitioned among 1,024 partitions using 3 methods: \myNum{i} hashing on subjects, \myNum{ii} hashing on objects, and \myNum{iii} random partitioning. Table \ref{table:part_methods}, shows statistics about the triples distribution among partitions for each method. Hashing on objects results in severely imbalanced partitions, whereas random partitioning and hashing on the subjects result in balanced partitions. We do not use random partitioning because it destroys data locality.

\nsstitle{Statistics Manager. }It maintains statistics about the RDF graph, which are used for global query planning and during adaptivity. Statistics are collected in a distributed manner during bootstrapping (Section \ref{sec:stat_collection}).

\nsstitle{Redistribution Controller. }It monitors the workload in the form of heat maps and triggers the adaptive Incremental ReDistribution (IRD) process for hot patterns. Only data accessed by hot patterns are redistributed and potentially replicated among workers. A redistributed hot pattern can be answered by all workers in parallel without communication. Using hierarchical representation, replicated hot patterns are indexed in a structure called Pattern Index (PI). Patterns in the PI can be combined for evaluating future queries without communication. Further, the controller implements replica replacement policy to keep replication within a threshold (Section \ref{sec:adaptivity}). 

\nsstitle{Locality-Aware Query Planner. }
Our planner uses the global statistics from the statistics manager and the pattern index from the redistribution controller to decide if a query, in whole or partially, can be processed without communication. Queries that can be fully answered without communication are planned and executed by each worker independently. On the other hand, for queries that require communication, the planner exploits the hash-based data locality and the query structure to find a plan that minimizes communication and the number of distributed joins (Section \ref{sec:distributed_query}).

\begin{table}\scriptsize
\centering
  {
  \caption{Triple distribution (in thousands of triples)}
  \begin{tabular}[h]{l|r|r|r||r|r|r}
      \hline
      &\multicolumn{3}{c||}{\textbf{LUBM-4000}} &\multicolumn{3}{c}{\textbf{YAGO2}}\\
      \hline
      Method& Max& Min& StDev& Max & Min & StDev\\ 
      \hline
      \textbf{hash(subj)} & 527 & 515 & 3& 296 & 267 & 3\\
      \textbf{hash(obj)} & 32,648 & 397 & 1,463& 9,914 & 140 & 663\\
      \textbf{random}  & 524 & 519 & 1& 280 & 276 & 1\\
      \hline
  \end{tabular}
  \label{table:part_methods}
  }
\end{table}

\nsstitle{Failure Recovery.} 
The master does not store any data but can be considered as a single-point of failure because it maintains the dictionaries, global statistics, and PI. A standard failure recovery mechanism (log-based recovery \cite{recovery1}) can be employed by {\logo}. Assuming stable storage, the master can recover by loading the dictionaries and global statistics because they are read-only and do not change in the system. The PI can be easily recovered by reading the query log and reconstructing the heat map.
Workers on the other hand store data; hence, in case of a failure, data partitions need to be recovered. Shen et al. \cite{recovery2} proposes a fast failure recovery solution for distributed graph processing systems. The solution is a hybrid of checkpoint-based and log-based recovery schemes. This approach can be used by {\logo} to recover worker partitions and reconstruct the replica index. However, reliability is outside this paper scope and we leave it for future work.

\vspace{-6pt}
\subsection{Worker}



\nsstitle{Storage Module.} Each worker $w_i$ stores its local set of triples $D_{i}$
in an in-memory data structure, which supports the following search operations, where $s$, $p$, and $o$ are subject, predicate, and object:
\begin{enumerate}
 \item given $p$, return set \mbox{$\{(s,o) \mid \langle s,p,o\rangle \in D_i\}$.}
 \item given $s$ and $p$, return set $\{o \mid \langle s,p,o\rangle \in D_i\}$.
 \item given $o$ and $p$, return set $\{s \mid \langle s,p,o\rangle \in D_i\}$.
\end{enumerate}
Since all the above searches require a known predicate, we primarily hash 
triples in each worker by predicate.
The resulting {\em predicate} index (simply P-index) immediately supports search by predicate (i.e., the first operation).
Furthermore, we use two hash maps to re-partition each bucket of triples having the same predicate, based on their subjects and objects, respectively.
These two hash maps support the second and third search operation
and they are called {\em predicate-subject} index 
(PS-index)
and {\em predicate-object} index (PO-index),
respectively.
Given the number of unique predicates is typically small, our storage scheme avoids unnecessary repetitions of predicate values. Note that when answering a query, if the predicate itself is a variable, then we simply iterate over all predicates. 
Our indexing scheme is tailored for typical RDF knowledge bases and their workloads, 
being orthogonal to the rest of the system (i.e., alternative schemes, like indexing all SPO combinations \cite{Neumann2010} could be used at each worker).
Finally, the storage module computes statistics about its local data and shares them with the master after data loading. 

\nsstitle{Replica Index.} Each worker has an in-memory \emph{replica index} that stores and indexes replicated data as a result of the adaptivity. This index initially contains no data and is updated dynamically by the incremental redistribution (IRD) process (Section~\ref{sec:adaptivity}).

\nsstitle{Query Processor. }Each worker has a query processor that operates in two modes: \myNum{i} \emph{Distributed Mode} for queries that require communication. In this case, all workers solve the query concurrently and exchange intermediate results (Section \ref{sec:query_eval}). 
\myNum{ii} \emph{Parallel Mode} for queries that can be answered without communication. Each worker has all the data needed for query evaluation locally (Section \ref{sec:adaptivity}).

\nsstitle{Local Query Planner. }Queries executed in parallel mode are planned by workers autonomously. 
For example, star queries joining on the subject are processed in parallel due to the initial partitioning. 
Moreover, queries answered in parallel after the adaptivity process are also planned by local query planners. 

\vspace{-5pt}
\subsection{Statistics Collection} 
\label{sec:stat_collection}
{\logo} collects and aggregates statistics from workers for global query planning and during the adaptivity process. Keeping 
statistics about each vertex in the entire RDF data graph is prohibitively expensive. {\logo} solves the problem by focusing on predicates rather than vertices. Therefore, the storage complexity of  statistics is linear to the number of unique predicates, which is typically very small compared to the data size. For each unique predicate $p$, we calculate the following statistics: 
\myNum{i} The \emph{cardinality} of $p$, denoted as $|p|$, is the number of triples in the data graph that have $p$ as predicate. 
\myNum{ii} $|p.s|$ and $|p.o|$ are the numbers of {\em unique subjects and objects} using predicate $p$, respectively. 
\myNum{iii} The \emph{subject} score of $p$, denoted as $\overline{p_S}$, is the average degree of all vertices $s$, such that $\RDF{s}{p}{\emph{?x}}\in D$. 
\myNum{iv} The \emph{object} score of $p$, denoted as $\overline{p_O}$, is the average degree of all vertices $o$, such that $\RDF{\emph{?x}}{p}{o}\in D$. 
\myNum{v} \emph{Predicates Per Subject} $P_{ps}$ = $|p|/|p.s|$  is the average number of triples with predicate $p$ per unique subject. 
\myNum{vi} \emph{Predicates Per Object} $P_{po}$ = $|p|/|p.o|$ is the average number of triples with predicate $p$ per unique object. 

\begin{figure}
  \centering
  \includegraphics[width=0.56\columnwidth]{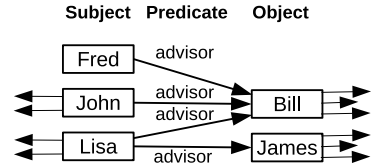}
  \caption{Statistics calculation for $p{=}advisor$, based on Figure~\ref{fig:rdf_graph}.}
  \label{fig:stat_calc}
\end{figure}

For example, Figure \ref{fig:stat_calc} illustrates the computed statistics for predicate {\em advisor} using the data graph of Figure \ref{fig:rdf_graph}. Since {\em advisor} appears four times with three unique subjects and two unique objects, $|p|=4$, $|p.s|=3$ and $|p.o|=2$. The subject score $\overline{p_S}$ is $(1+3+4)/3=2.67$ because {\em advisor} appears with four unique subjects: Fred, John and Lisa, whose degrees (i.e., in-degree plus out-degree) are 1, 3 and 4, respectively. Similarly, $\overline{p_O}=(6+4)/2=5$. Finally, the number of predicates per subject $P_{ps}$ is $4/3=1.3$ because Lisa is associated with two instances of the predicate (i.e., two advisors).


\vspace{-5pt}
\subsection{System overview}

Here we give an abstract overview of {\logo}. After encoding and partitioning the data, each worker loads its triples and collects local statistics. The master node aggregates these statistics and {\logo} starts answering queries. A user submits a SPARQL query $Q$ to the master. The query planner at the master consults the redistribution controller to decide whether $Q$ can be executed in parallel mode. 
The redistribution controller uses global statistics to transform $Q$ into a hierarchical representation $Q'$ (details in Section \ref{sec:query_decomposition}). If $Q'$ exists in the Pattern Index (PI) or if $Q'$ is a star query joining on the subject column, then $Q$ can be answered in parallel mode; otherwise, it is executed in distributed mode. 
If $Q$ is executed in distributed mode, the locality-aware planner devises a global query plan. Each worker gets a copy of this plan and evaluates the query accordingly. If $Q$ can be answered in parallel mode, the master broadcasts the query to all workers. Each worker generates its local query plan using local statistics and executes $Q$ without communication.

As more queries get submitted to the system, the redistribution controller updates the heat map, identifies hot patterns, and triggers the IRD process. Consequently, {\logo} adapts to the query load by answering more queries in parallel mode.

%% file: sec_distributed.tex
\section{Query Evaluation}
\label{sec:distributed_query}


A basic SPARQL query consists of multiple subquery triple patterns: $q_1, q_2, \ldots , q_n$. Each subquery includes variables or constants, some of which are used to bind the patterns together, forming the entire query graph (e.g., see Figure \ref{fig:queryProf}(b)).
A query with $n$ subqueries requires the evaluation of $n-1$ joins. 
Since data are memory resident and hash-indexed, we favor hash joins as they prove to be competitive to more sophisticated join methods \cite{Blanas}. Our query planner devises an ordering of these subqueries and generates a left-deep join tree, where the right
operand of each join is a base subquery (not an intermediate result). We do not use bushy tree plans to avoid building indices for intermediate results. 


\begin{figure*}
  \centering
  \subfigure[$q_1$, $q_2$, $q_3$]{
  	\label{fig:plan1}
  	\includegraphics[width=1\columnwidth]{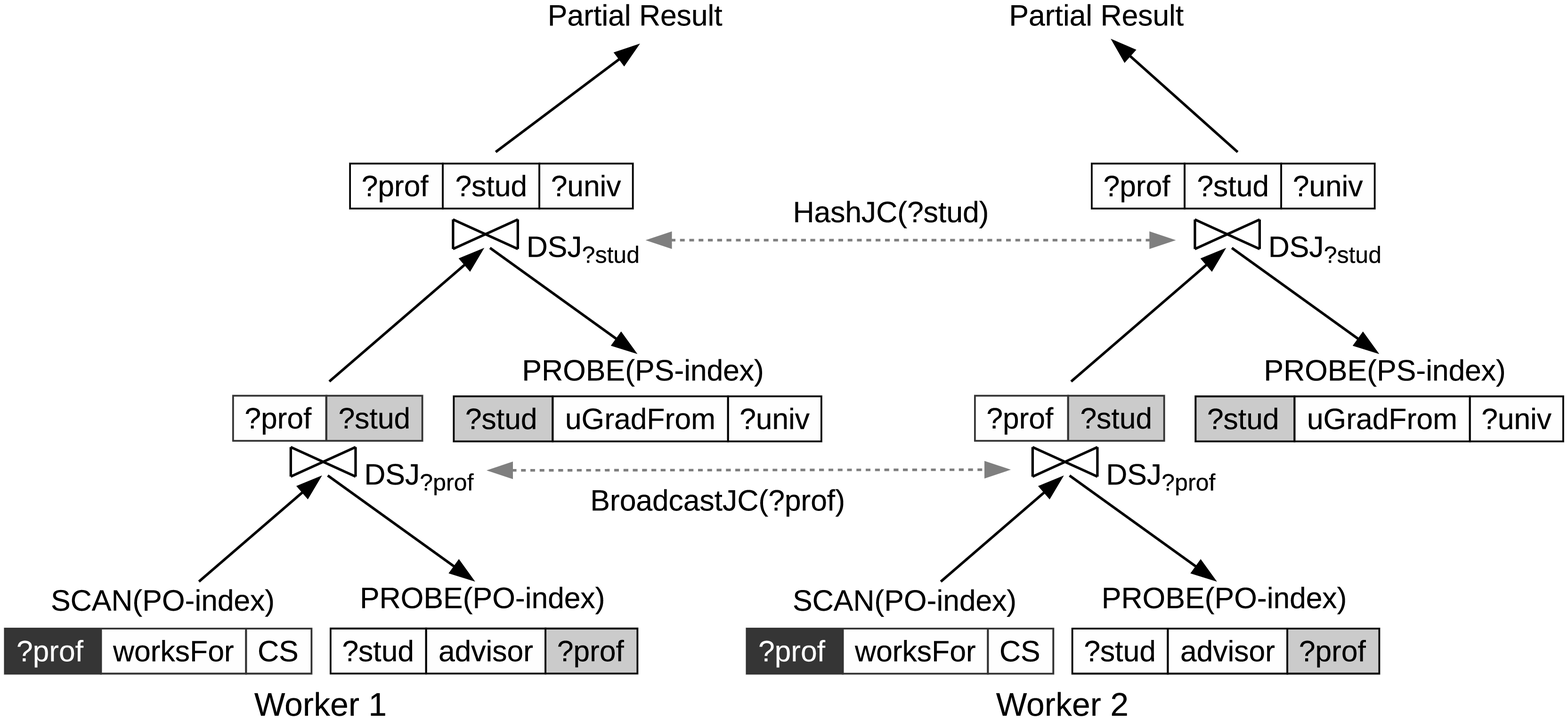}
  }
  \subfigure[$q_2$, $q_1$, $q_3$]{
  	\label{fig:plan2}
  	\includegraphics[width=1\columnwidth]{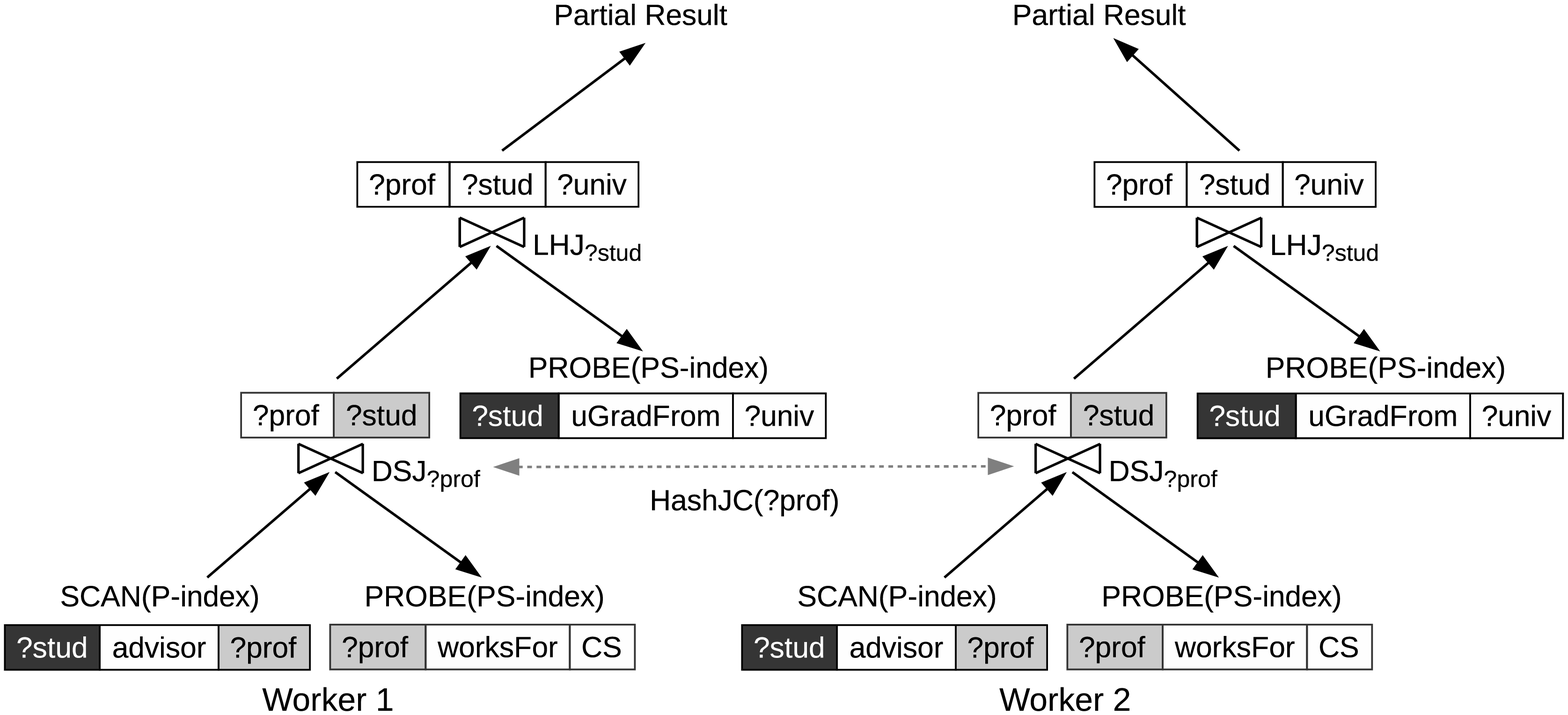}
  }
  \caption{Executing query $Q_{prof}$ using two different subquery orderings.}
\end{figure*}

\subsection{Distributed Query Evaluation}
\label{sec:query_eval}

In {\logo}, triples are hash partitioned among many workers based on subject values. Consequently, subject star queries (i.e. all subqueries join on the subject column) can be evaluated locally in parallel without communication. However, for other types of queries, workers may have to communicate intermediate results during join evaluation. For example, consider the query in Figure \ref{fig:queryProf} and the partitioned data graph in Figure \ref{fig:rdf_graph}. The query consists of two subqueries $q_1$ and $q_2$, where:
\begin{itemize} 
    \item $q_1$: \RDF{?prof}{worksFor}{CS}
    \item $q_2$: \RDF{?stud}{advisor}{?prof}  
\end{itemize}

\begin{table}
  \centering
  \caption{Matching result of $q_1$ on workers $w_1$ and $w_2$.}
  \begin{tabular}{cllccll}
      \multicolumn{1}{l}{} & $w_1$ &  & \multicolumn{1}{l}{}  & \multicolumn{1}{l}{} & $w_2$ &  \\ \cline{1-3} \cline{5-7} 
      \multicolumn{3}{|c|}{\cellcolor[HTML]{9B9B9B}?prof}     & \multicolumn{1}{c|}{} & \multicolumn{3}{c|}{\cellcolor[HTML]{9B9B9B}?prof}      \\ \cline{1-3} \cline{5-7} 
      \multicolumn{3}{|c|}{\texttt{James}}  & \multicolumn{1}{c|}{} & \multicolumn{3}{c|}{\texttt{Bill}}    \\ \cline{1-3} \cline{5-7} 
  \end{tabular}
  \label{tab:q1_res}
\end{table}

\begin{table}
\centering
\caption{The final query results $q_1 \bowtie q_2$ on both workers.}
\begin{tabular}{ccccc}
\multicolumn{2}{c}{$w_1$}                                                                                   &                       & \multicolumn{2}{c}{$w_2$}                                                                                  \\ \cline{1-2} \cline{4-5} 
\multicolumn{1}{|c|}{\cellcolor[HTML]{9B9B9B}?prof} & \multicolumn{1}{c|}{\cellcolor[HTML]{9B9B9B}?stud} & \multicolumn{1}{c|}{} & \multicolumn{1}{c|}{\cellcolor[HTML]{9B9B9B}?prof} & \multicolumn{1}{c|}{\cellcolor[HTML]{9B9B9B}?stud} \\ \cline{1-2} \cline{4-5} 
\multicolumn{1}{|c|}{James}                         & \multicolumn{1}{c|}{Lisa}                          & \multicolumn{1}{c|}{} & \multicolumn{1}{c|}{Bill}                          & \multicolumn{1}{c|}{Lisa}                          \\ \cline{1-2} \cline{4-5} 
                                                    &                                                    & \multicolumn{1}{c|}{} & \multicolumn{1}{c|}{Bill}                          & \multicolumn{1}{c|}{John}                          \\ \cline{4-5}
                                                    &                                                    & \multicolumn{1}{c|}{} & \multicolumn{1}{c|}{Bill}                          & \multicolumn{1}{c|}{Fred}                          \\ \cline{4-5} 
\end{tabular}
\label{tab:final_res}
\end{table}

The query is evaluated by a single subject-object join; however, neither of the workers has all the data needed for evaluating the entire query. In other words, workers need to communicate because objects' locality is not known. To solve such queries, {\logo} employs the Distributed Semi-Join (DSJ) algorithm. Each worker scans the PO-index to find all triples matching $q_1$. The results on workers $w_1$ and $w_2$ are shown in Table \ref{tab:q1_res}. 
Then, each worker creates a projection on the join column $?prof$ and exchanges it with the other worker. Once the projected column is received, each worker computes the semi-join $q_1 \rtimes_{?prof} q_2$ using its PO-index. Specifically, $w_1$ probes $p=\texttt{advisor}, o=\texttt{Bill}$ while $w_2$ probes $p=\texttt{advisor}, o=\texttt{James}$ to their PO-index. 
Note that workers also need to evaluate semi-joins using their local projected column. Then, the semi-join results are shipped to the sender. In this case, $w_1$ sends \RDF{Lisa}{advisor}{Bill} and \RDF{Fred}{advisor}{Bill} to $w_2$; no candidate triples are sent from $w_2$ because \texttt{James} has no advisees on $w_2$. Finally, each worker computes the final join $q_1 \bowtie_{?prof} q_2$. The final query results at both workers are shown in Table \ref{tab:final_res}.

\subsubsection{Hash-based data locality}

\begin{myobs}
\label{obs:locality}
DSJ can benefit from subject hash locality to minimize communication. If the join column of the right operand is subject, the projected column of the left operand is  hash distributed by all workers. Otherwise, the projected column on each worker is broadcasted to all other workers.
\end{myobs}

In our example, since the join column of $q_2$ is the object column ($?prof$), each worker sends the entire join column to the other worker. However, based on Observation \ref{obs:locality}, communication can be minimized if the join order is reversed (i.e., $q_2 \bowtie q_1$). In this case, each worker scans the P-index to find triples matching $q_2$ and creates a projection on $?prof$. Then, because $?prof$ is the subject of $q_1$, both workers exploit the subject hash-based locality by partitioning the projection column and communicating each partition to the respective worker,
as opposed to broadcasting the entire projection column to all workers. 
Consequently, $w_1$ sends \texttt{Bill} to only $w_2$ because of Bill's hash value. The final query results are shown in Table \ref{tab:final_res2}. Notice that the final results are the same for both query plans; however, the results reported by each worker are different.

\begin{table}
\centering
\caption{The final query results $q_2 \bowtie q_1$ on both workers.}
\begin{tabular}{ccccc}
\multicolumn{2}{c}{$w_1$}                                                                                   &                       & \multicolumn{2}{c}{$w_2$}                                                                                  \\ \cline{1-2} \cline{4-5} 
\multicolumn{1}{|c|}{\cellcolor[HTML]{9B9B9B}?prof} & \multicolumn{1}{c|}{\cellcolor[HTML]{9B9B9B}?stud} & \multicolumn{1}{c|}{} & \multicolumn{1}{c|}{\cellcolor[HTML]{9B9B9B}?prof} & \multicolumn{1}{c|}{\cellcolor[HTML]{9B9B9B}?stud} \\ \cline{1-2} \cline{4-5} 
\multicolumn{1}{|c|}{James}                         & \multicolumn{1}{c|}{Lisa}                          & \multicolumn{1}{c|}{} & \multicolumn{1}{c|}{Bill}                          & \multicolumn{1}{c|}{John}                          \\ \cline{1-2} \cline{4-5} 
\multicolumn{1}{|c|}{Bill}                          & \multicolumn{1}{c|}{Lisa}                          &                       &                                                    &                                                    \\ \cline{1-2}
\multicolumn{1}{|c|}{Bill}                          & \multicolumn{1}{c|}{Fred}                          &                       &                                                    &                                                    \\ \cline{1-2}
\end{tabular}
\label{tab:final_res2}
\end{table}

\subsubsection{Pinned subject}

\begin{myobs}
\label{obs:pinned}
Under the subject hash partitioning, combining right-deep tree planning and the DSJ algorithm for solving SPARQL queries, causes the intermediate and final results to be local to the subject of the first executed subquery pattern $p_1$. We refer to this subject as \texttt{pinned\_subject}.
\end{myobs}

In our example, executing $q_1$ first causes $?prof$ to be the \emph{pinned\_subject} because it is the subject of $q_1$. Hence, the intermediate and final results are local (pinned) to the bindings of $?prof$,  \texttt{James} and \texttt{Bill} in $w_1$ and $w_2$, respectively. Changing the order by executing $q_2$ first made $?stud$ to be the \emph{pinned\_subject}. Accordingly, the results are pinned at the bindings of $?stud$.

Consequently, {\logo} leverages Observations \ref{obs:locality} and \ref{obs:pinned} to minimize communication and synchronization overhead. To see this, consider $Q_{prof}$ which extends the query in Figure \ref{fig:queryProf} with one more triple pattern, namely $q_3$: \RDF{?stud}{uGradFrom}{?univ}. Assume $Q_{prof}$ is executed in the following order: $q_1$, $q_2$, $q_3$. The query execution plan is pictorially shown in Figure \ref{fig:plan1}. The results of the first join (i.e., $q_1 \bowtie q_2$) is shown in Table \ref{tab:final_res} where $?prof$ is the \emph{pinned\_subject} as demonstrated above. The query continues by joining the intermediate result ($q_1 \bowtie q_2$) with $q_3$ on $?stud$, the subject of $q_3$. Both workers projects the intermediate results on $?stud$ and hash distribute the bindings of $?stud$ (Observation \ref{obs:locality}). Then, all workers evaluate semi-joins with $q_3$ and return the candidate triples to the other workers where the final query results are formulated.

Notice that the execution order $q_1$, $q_2$, $q_3$ requires communication for evaluating both joins. 
Nonetheless, a better ordering that would potentially minimize communication is $q_2$, $q_1$, $q_3$. The execution plan is shown in Figure \ref{fig:plan2}. The first join (i.e.,  $q_2 \bowtie q_1$) already proved to incur less communication by avoiding the need for broadcasting the entire projection column. The results of this join is pinned at $?stud$ as shown in Table \ref{tab:final_res2}. Since the join column of $q_3$ ($?stud$) is the $pinned\_subject$, joining the intermediate results ($q_2 \bowtie q_1$) with $q_3$ can be processed locally by each worker without communication using Local Hash Join (LHJ). Therefore, the ordering of the subqueries affects the amount of communication incurred during query execution.

\subsubsection{The four cases of a join}
 Formally, joining two subqueries, say $p_i$ (possibly an intermediate pattern) and $p_{j}$, has four possible scenarios: 
the first three assume that $p_i$ and $p_{j}$ join on columns $c_1$ and $c_2$, respectively. 
\myNum{i} If \emph{$c_2$ = subject} AND $c_2 =$ \emph{pinned\_subject}, then the join can be answered by all workers in parallel without communication. 
\myNum{ii} If \emph{$c_2$ = subject} AND $c_2 \neq$ \emph{pinned\_subject}, then the join is 
evaluated using DSJ; but the projected join column of $p_i$ is hash distributed. 
\myNum{iii} If $c_2 \neq$ \emph{subject}, then the join is executed using DSJ and the projected join column of $p_i$ is sent from all workers to all other workers. This includes joining on the object or predicate column.
Finally, \myNum{iv} if $p_i$ and $p_{j}$ join on multiple columns, we opt to join on the subject column of $p_{j}$, if it is a join attribute. 
This allows the join column of $p_i$ to be hash distributed as in \emph{(ii)}.
If the subject column of $p_{j}$ is not a join attribute, we join on another column of $p_{j}$ and  
broadcasting the projection column to all workers, as in scenario \emph{(iii)}. Verifying on the other columns is carried out during the join finalization by the DSJ.

\input{distrib_algo.tex}

\subsubsection{Evaluation of join orderings}
 
Based on the above four scenarios, we introduce our Locality-Aware Distributed Query Execution algorithm (see Algorithm \ref{algo:distrib_query_algorithm}). The algorithm receives an ordering of the subquery 
patterns. For each join iteration, if the second subquery joins on a subject which is the pinned subject, the join is executed without communication (line 7). 
Otherwise, the join is evaluated with the DSJ algorithm (lines 9-28). 
In the first iteration, $p_1$ is a base subquery pattern; however, for the subsequent iterations $p_1$ is a pattern of intermediate results. If $p_1$ is the first subquery to be matched, each worker finds the local matching of $p_1$ (line 2) and projects on the join column $c_1$ (line 5). If the join column of $q_2$ is subject, then each worker hash distributes the projected column (line 7); or sends it to all other workers otherwise (line 9). All workers perform semi-join on the received data (line 14) and send the results back to $w$ (line 15). Finally, each worker finalizes the join (line 19) and formulates the final result (line 20). Lines 14 and 19 are implemented as local hash-joins using the local index in each worker. The final result of a DSJ iteration becomes $p_1$ in the next iteration.

Algorithm \ref{algo:distrib_query_algorithm} can solve star queries that join on the subject in parallel mode. However, the planning is done by the master using global statistics. We argue that allowing each worker to plan the query execution autonomously would result in a better performance. For example, using the data graph in Figure \ref{fig:rdf_graph}, Table \ref{tab:matching_star} shows triples that match the following star query:

\begin{itemize} 
    \item $q_1$: \RDF{?s}{advisor}{?p}
    \item $q_2$: \RDF{?s}{uGradFrom}{?u} 
\end{itemize}

Any global plan (i.e., $q_1 \bowtie q_2$ or $q_2 \bowtie q_1$) would require a total of four index lookups to solve the join. However, $w_1$ and $w_2$ can evaluate the join using 2 and 1 index lookup(s), respectively. Therefore, to solve such queries, the master sends the query to all workers; each worker utilizes its local statistics to formulate the execution plan, evaluates the query locally without communication, and sends the final result to the master. 

\begin{table}
\caption{Triples matching \RDF{?s}{advisor}{?p} and \RDF{?s}{uGradFrom}{?u} on two workers.}
\resizebox{\columnwidth}{!}{%
\begin{tabular}{cccc|lccc}
\multicolumn{3}{c}{Worker 1}                                                                                                                                & \multicolumn{1}{l|}{} &                       & \multicolumn{3}{c}{Worker 2}                                                                                                                               \\ \cline{1-3} \cline{6-8} 
\multicolumn{1}{|c|}{\cellcolor[HTML]{9B9B9B}advisor}   & \multicolumn{1}{c|}{\cellcolor[HTML]{9B9B9B}?s} & \multicolumn{1}{c|}{\cellcolor[HTML]{9B9B9B}?p} &                       & \multicolumn{1}{l|}{} & \multicolumn{1}{c|}{\cellcolor[HTML]{9B9B9B}advisor}   & \multicolumn{1}{c|}{\cellcolor[HTML]{9B9B9B}?s} & \multicolumn{1}{c|}{\cellcolor[HTML]{9B9B9B}?p} \\ \cline{1-3} \cline{6-8} 
\multicolumn{1}{c|}{}                                   & \multicolumn{1}{c|}{Fred}                       & \multicolumn{1}{c|}{Bill}                       &                       &                       & \multicolumn{1}{c|}{}                                  & \multicolumn{1}{c|}{John}                       & \multicolumn{1}{c|}{Bill}                       \\ \cline{2-3} \cline{7-8} 
\multicolumn{1}{c|}{}                                   & \multicolumn{1}{c|}{Lisa}                       & \multicolumn{1}{c|}{Bill}                       &                       &                       &                                                        &                                                 &                                                 \\ \cline{2-3}
\multicolumn{1}{c|}{}                                   & \multicolumn{1}{c|}{Lisa}                       & \multicolumn{1}{c|}{James}                      &                       &                       &                                                        &                                                 &                                                 \\ \cline{2-3}
                                                        &                                                 &                                                 &                       &                       &                                                        &                                                 &                                                 \\ \cline{1-3} \cline{6-8} 
\multicolumn{1}{|c|}{\cellcolor[HTML]{9B9B9B}uGradFrom} & \multicolumn{1}{c|}{\cellcolor[HTML]{9B9B9B}?s} & \multicolumn{1}{c|}{\cellcolor[HTML]{9B9B9B}?u} &                       & \multicolumn{1}{l|}{} & \multicolumn{1}{c|}{\cellcolor[HTML]{9B9B9B}uGradFrom} & \multicolumn{1}{c|}{\cellcolor[HTML]{9B9B9B}?s} & \multicolumn{1}{c|}{\cellcolor[HTML]{9B9B9B}?u} \\ \cline{1-3} \cline{6-8} 
\multicolumn{1}{c|}{}                                   & \multicolumn{1}{c|}{Lisa}                       & \multicolumn{1}{c|}{MIT}                        &                       &                       & \multicolumn{1}{c|}{}                                  & \multicolumn{1}{c|}{Bill}                       & \multicolumn{1}{c|}{CMU}                        \\ \cline{2-3} \cline{7-8} 
\multicolumn{1}{c|}{}                                   & \multicolumn{1}{c|}{James}                      & \multicolumn{1}{c|}{CMU}                        &                       &                       & \multicolumn{1}{c|}{}                                  & \multicolumn{1}{c|}{John}                       & \multicolumn{1}{c|}{CMU}                        \\ \cline{2-3} \cline{7-8} 
\end{tabular}
}
\label{tab:matching_star}
\end{table}

\subsection{Locality-Aware Query Optimization}
\label{sec:planner}
Our locality-aware planner leverages the query structure and the hash-based data distribution during query plan generation to minimize communication. Accordingly, the planner uses a cost-based optimizer for finding the best subqueries ordering. We use Dynamic Programming (DP) for optimizing the plan. 

Each state $S$ in DP is identified by a subgraph $\varrho$ of the query graph. A state can be reached by different orderings on $\varrho$. Therefore, we maintain in each state the ordering that results in the least estimated communication cost ($S.cost$). We also keep estimated cardinalities of the variables in the query. Furthermore, instead of maintaining the cardinality of the state, we keep the cumulative cardinality of all intermediate results that led to this state. The reason is that the cardinality of the state will be the same regardless of the ordering. However, reaching to the same state using different ordering will result in different cumulative cardinality. 

We initialize a state $S$ for each subquery pattern (subgraph of size 1) $p_i$. 
$S.cost$ is initially zero because a query with a single pattern can be answered without communication. Then, we expand the subgraph by joining with another pattern $p_j$, leading to a new state $S'$ such that:
\[ 
S'\hspace{-2pt}.cost = min(S'\hspace{-2pt}.cost,\, S.cost+cost(S, p_j))
\]

If we reach a state using different orderings with the same cost, we keep the one with the least cumulative cardinality. This happens for subqueries that join on the $pinned\_subject$. To minimize the DP table size, we maintain a global minimum cost ($minC$) of all found plans. Because our cost function is monotonically increasing, any branch that results in a cost $> minC$ is pruned. Moreover, because of Observation 1, we start the DP process by considering subqueries connected to the subject with the highest number of outgoing edges. Considering these subqueries first increases the probability of converging to the optimal plan faster. 

\subsection{Cost Estimation}
\label{sec:cost}
We set the initial communication cost of DP states to zero. Cardinalities of subqueries with variable subjects and objects are already captured in the master's global statistics. Hence, we set the cumulative cardinalities of the initial states to the cardinalities of the subqueries themselves and set the size of the subject and object bindings to $|p.s|$ and $|p.o|$. 
Furthermore, the master consults the workers to update the cardinalities of subquery patterns that are attached to constants or have unbounded predicates. This is done locally at each worker by simple lookups to its PS- and PO- indices to update the cardinalities of variables bindings \mbox{accordingly.}

We estimate the cost of expanding a state $S$ with a subquery $p_j$, where $c_j$ and $P$ are the join column and the predicate of $p_j$, respectively. If the join does not incur communication, the cost of the new state $S'$ is zero. Otherwise, the expansion is carried out through DSJ and we incur two phases of communication: \myNum{i} transmitting the projected join column and \myNum{ii} replying with the candidate triples.  Estimating the communication in the first phase depends on the cardinality of the join column bindings in $S$, denoted as $B(c_j)$. In the second phase, communication depends on the selectivity of the semi-join and the number of variables $\nu$ in $p_j$ (constants are not communicated). Moreover, if $c_j$ is the subject column of $p_j$, we hash distribute the projected column. Otherwise, the column needs to be sent to all workers. The cost of expanding $S$ with $p_j$ is:

{\small
\[ cost(S, p_j) = \left\{ 
  \begin{array}{l l}
   0 & \\
   \text{if $c_j$ is subject \& $c_j=pinned\_subject$}\\
   \\
   S.B(c_j)+(\nu \cdot S.B(c_j) \cdot P_{ps}) &\\
   \text{if $c_j$ is subject \& $c_j \neq pinned\_subject$}\\
   \\
   (S.B(c_j) \cdot N)+(\nu \cdot N \cdot S.B(c_j) \cdot P_{po}) &\\
   \text{if $c_j$ is not subject}
  \end{array} \right.\]
}
  
Next, we need to re-estimate the cardinalities of all variables in $p_j$. For each variable $\overline{v} \in p_j$, let $|p.\overline{v}|$ denote $|p.s|$ or $|p.o|$ if $\overline{v}$ is subject or object, respectively. Similarly, let $P_{p\overline{v}}$ denote $|P_{ps}|$ if $\overline{v}$ is subject or $|P_{po}|$ if $\overline{v}$ is object. We re-estimate the cardinality of $\overline{v}$ in the new state $S'$ as:

{\small
\[ S'.B(\overline{v}) = \left\{ 
  \begin{array}{l l}
   min(S.B(\overline{v}), |P|) &  \text{if $\nu=1$}\\
   min(S.B(\overline{v}), |p.\overline{v}|) &  \text{if $\overline{v}=c_j$ \& $\nu>1$}\\
   min(S.B(\overline{v}), S.B(\overline{v}) \cdot P_{p\overline{v}} &\\
   ~~~~~~, |p.\overline{v}|) &  \text{if $\overline{v} \neq c_j$ \& $\nu>1$}\\
\end{array} \right.\]
}

We use the cumulative cardinality when we reach the same state using two different orderings. Therefore, we also re-estimate the cumulative state cardinality $|S'|$. Let $P_{pc_j}$ denote $|P_{ps}|$ or $|P_{po}|$ depending on the position of $c_j$, $|S'| = |S| \cdot (1+P_{pc_j}) $.
Notice that we use an upper bound estimation of the cardinalities. A special case of the last equation is when a subquery has a constant. In this case, we assume that each tuple in the previous state has a connection to this constant by setting $P_{pc_j}$ to 1. 


%% file: distrib_algo.tex
\begin{algorithm}[t]
 	\scriptsize
	\KwIn{Query $Q$ with $n$ ordered subqueries $\{q_1, q_2, \ldots q_n\}$}
	\KwResult{Answer of $Q$}
	\SetKwFunction{answerSub}{answerSubquery}
	\SetKwFunction{getjc}{getJoinColumns}
	\SetKw{Let}{Let}
	\SetKwComment{cc}{// }{}
	\SetKwFunction{nocomm}{JoinWithoutCommunication}
	
	$p_1 \gets$ $q_1$\;
	$pinned\_subject \gets$ $p_1.subject$\;
	\For{$i\leftarrow 2$ \KwTo $n$}{
		$p_2 \gets$  $q_i$\;
		[$c_1$, $c_2$]$\gets$ \getjc{$p_1$, $p_2$}\;
		\If{$c_2$ == $pinned\_subject$ AND $c_2$ is $subject$}{
			$p_1 \gets$ \nocomm($p_1$, $p_2$, $c_1$, $c_2$);\
		}
		\Else{
			\If{$p_1$ NOT intermediate pattern}{
				$RS_1 \gets$  \answerSub{$p_1$}\;
			}
			\Else{
				$RS_1$ is the result of the previous join 
			}
			$RS_1[c_1] \gets$ $\pi_{c_1}(RS_1)$; \cc{projection on $c_1$}
			\If{$c_2$ is $subject$}{
				Hash $RS_1[c_1]$  among workers\;
			}
			\Else{
				Send $RS_1[c_1]$ to all workers\;
			}
			\Let {$RS_2 \gets$  \answerSub{$p_2$}}\;
			\ForEach{worker $w$, $w:$ $1 \rightarrow N$}{
				\Let{$RS_{1w}[c_1]$ denote the $RS_1[c_1]$ received from $w$}\\
				\Let{$CRS_{2w}$ be the candidate triples of $RS_2$ that join with $RS_{1w}[c_1]$}\\
				$CRS_{2w} \gets$ $RS_{1w}[c_1] \bowtie_{RS_{1w}[c_1].c_1 = RS_2.c_2} RS_2$\;
				Send $CRS_{2w}$ to worker $w$\;
			}
			\ForEach{worker $w$, $w:$ $1 \rightarrow N$}{
				\Let{$RS_{2w}$ be the $CRS_{2w}$ received from worker $w$}\\
				\Let{$RES_w$ be the result of joining with worker $w$}\\
				$RES_w \gets$ $RS_1 \bowtie_{RS_1.c_1 = RS_{2w}.c_2} RS_{2w}$\;
			}
			$p_1 \gets$ $RES_1 \cup RES_2 \cup .... \cup RES_N$\;
		}
	}		
	\caption{Locality-Aware Distributed Execution}
	\label{algo:distrib_query_algorithm}
\end{algorithm}

%% file: sec_adaptivity.tex
\vspace{-3pt}
\section{{\logo} Adaptivity}
\label{sec:adaptivity}

Studies show that even minimal communication results in significant performance degradation \cite{Huang11:scalable,shape}. Thus, data need to be redistributed to minimize, if not eliminate, communication and synchronization overheads. {\logo} adapts to workload by redistributing only the parts of data needed for the current workload and adapts as the workload changes. The incremental redistribution model of {\logo} is a combination of hash partitioning and $k$-hop replication; however, it is guided by the query load rather than the data itself. Specifically, given a hot pattern $Q$ (hot patterns detection is discussed in Section \ref{sec:monitoring}), our system selects a special vertex in the pattern called the $core$ vertex (Section \ref{sec:vertexScore}). The system groups the data accessed by the pattern around the bindings of this core vertex. To do so, the system transforms the pattern into a redistribution tree rooted at the core (Section \ref{sec:query_decomposition}). Then, starting from the core vertex, first hop triples are hash distributed based on the core bindings. Next, triples that bind to the second level subqueries are collocated and so on (Section \ref{sec:incremental_redistribution}). {\logo} utilizes these redistributed patterns to answer queries in parallel without communication. 

\vspace{-2pt}
\subsection{Core Vertex Selection}
\label{sec:vertexScore}

\begin{figure}
 \centering
  \subfigure[Core is $?univ$] {
  		\label{fig:propagateUni}
\includegraphics[width=0.4\columnwidth]{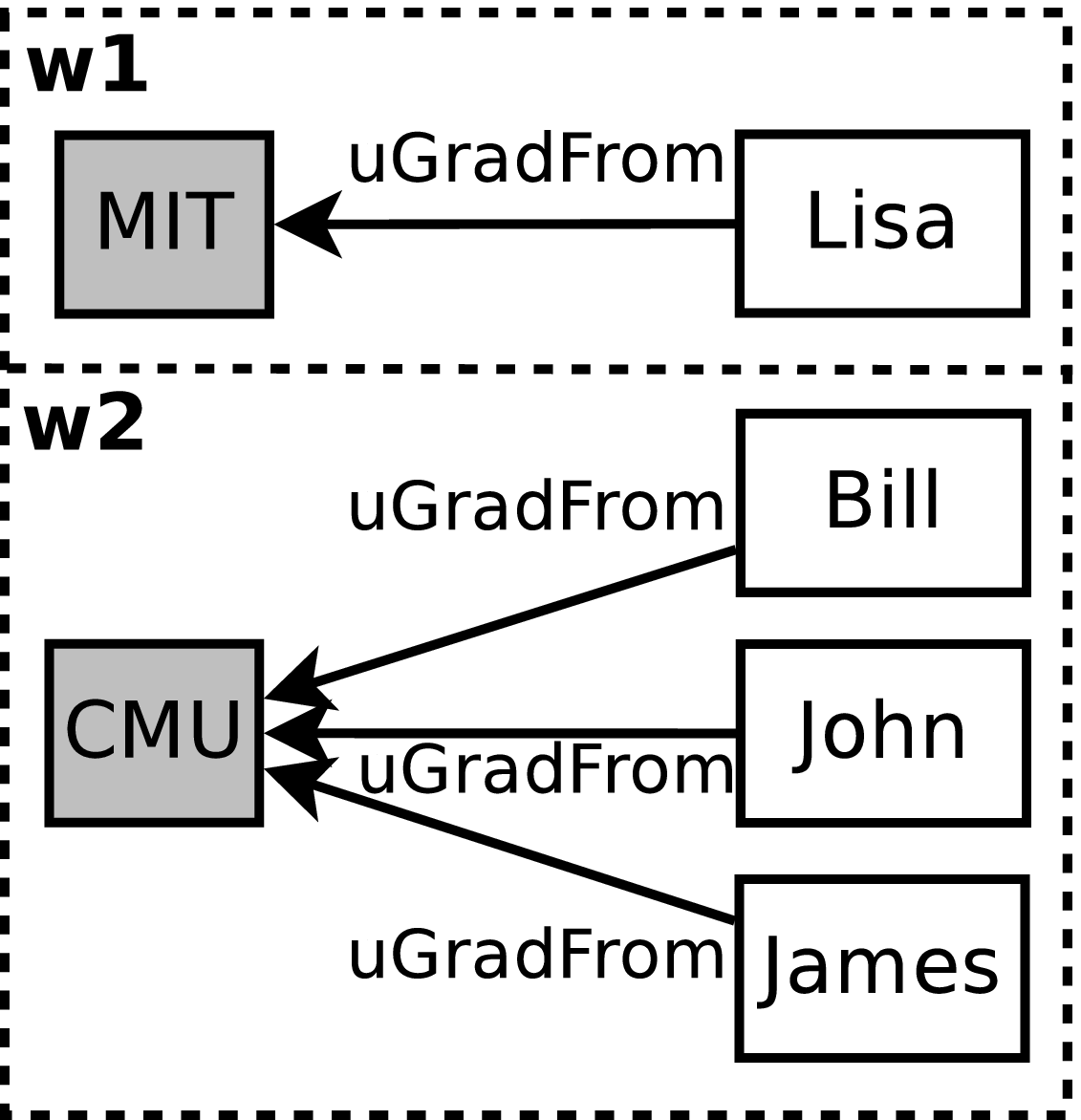}
  } \subfigure[Core is $?stud$]{
  		\label{fig:propagateStu}
\includegraphics[width=0.4\columnwidth]{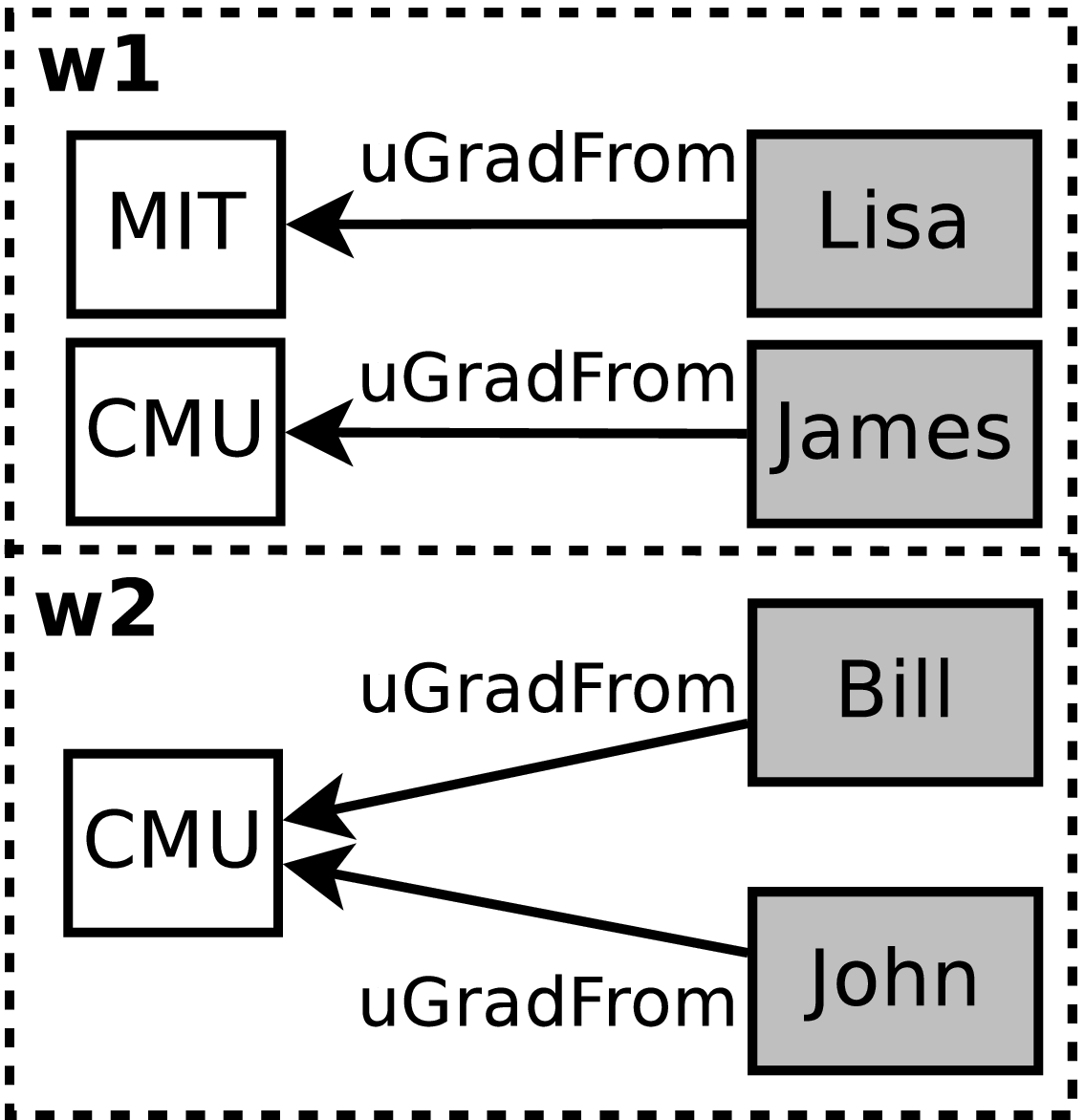}
  }
  \caption{Effect of choice of core on replication. In (a) there is no replication. In (b) CMU is both workers.}
\end{figure}

For a hot pattern, the choice of the core has a significant impact on the amount of replicated data as well as on query execution performance. 
For example, consider query $Q_1 =$
\RDF{?stud}{uGradFrom}{?univ}. Assume there are two workers, $w_1$ and $w_2$, and refer to the graph of Figure~\ref{fig:rdf_graph}; MIT and CMU are the bindings of $?univ$, whereas Lisa, John, James and Bill bind to $?stud$. 
Assume that $?univ$ is the core, then triples matching $Q_1$ will be hashed on the bindings of $?univ$ as shown in Figure \ref{fig:propagateUni}. Note that every binding of $?stud$ appears in one worker only. Now assume that $?stud$ is the core and triples are hashed using the bindings of $?stud$. This causes the binding $?univ$=CMU to exist on both workers (see Figure~\ref{fig:propagateStu}). The problem becomes more pronounced when the query has more triple patterns.
Consider $Q_2 = Q_1$ \texttt{AND} 
\RDF{?prof}{gradFrom}{?univ} and assume that $?stud$ is chosen as core. Because CMU exists on both workers, all its graduates will also be replicated (i.e., triples matching \RDF{?prof}{gradFrom}{CMU} will be replicated on both workers). Replication can become significant because it grows exponentially with the number of triple patterns \cite{Huang11:scalable}.

Intuitively, if random walks start from two random vertices (e.g., students), the probability of reaching the same well-connected vertex (e.g., university) within a few hops is higher than reaching the same student from two universities. In order to minimize replication, we must avoid reaching the same vertex when starting from the core. Hence, it is reasonable to select a well-connected vertex as the core. 

In the literature there are many definitions of what constitutes a well-connected vertex, many of which are based on complex data mining algorithms. In contrast, we employ a definition that poses minimal computational overhead. We assume that connectivity is proportional to degree centrality (i.e., in-degree plus out-degree edges) of a vertex. 
However, many RDF datasets follow the power-law distribution, where few vertices are of extremely high degrees. For example, vertices that appear as objects in triples with \emph{rdf:type} have very high degree centrality. Treating such vertices as cores results in imbalanced partitions and prevents the system from taking full advantage of 
parallelism \cite{Huang11:scalable}.

Recall from Section~\ref{sec:stat_collection} that we maintain statistics $\overline{p_S}$ and $\overline{p_O}$ for each predicate $p \in P$, where $P$ is the set of all predicates in the data. Let $P_s$ and $P_o$ be the set of all $\overline{p_S}$ and $\overline{p_O}$, respectively. We filter out predicates with extremely high scores and consider them outliers. Outliers are filtered out using Chauvenet's criterion \cite{outliars} on $P_s$ then $P_o$. If a predicate $p$ is detected as an outlier, we set: $\overline{p_S}= \overline{p_O} = -\infty$; else use $\overline{p_S}$ and $\overline{p_O}$ as computed in Section~\ref{sec:stat_collection}. Now, we can compute a score for each vertex in the query as follows:

\begin{mydef}[Vertex score]
For a query vertex $v$, let $E_{out}(v)$ be the set of outgoing edges and $E_{in}(v)$ be the set of incoming edges. Also, let $A$ be the set of all $\overline{p_S}$ for the $E_{out}(v)$ edges and all $\overline{p_O}$ for $E_{in}(v)$ edges. The vertex score $\overline{v}$ is defined as: $\overline{v}=\max(A)$.
\end{mydef}

\begin{figure}
	\centering 
	\includegraphics[width=0.5\linewidth]{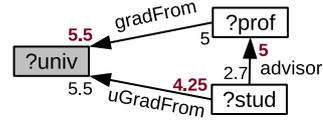}
	\caption{Example of vertex score: numbers correspond to $\overline{p_S}$ and $\overline{p_O}$ values.
			Assigned vertex scores $\overline{v}$ are shown in bold.}
	\label{fig:score_example}
\end{figure}

Figure~\ref{fig:score_example} shows an example for vertex score assignment. For vertex $?prof$,
$E_{in}(?prof)=$ \{\texttt{advisor}\} and $E_{out}(?prof)=$ \{\texttt{gradFrom}\}. Both predicates (i.e., advisor and gradFrom) contribute a score of 5 to $?prof$. Therefore, $\overline{?prof}=5$.

\begin{mydef}[Core vertex]
Given a query $Q$, the vertex $v'$ with the highest score is called the core vertex.
\end{mydef}

\noindent{In Figure \ref{fig:score_example}, $?univ$ has the highest score, hence, it is the core vertex for this pattern.}

\input{graph_walk.tex}

\subsection{Generating the Redistribution Tree}
\label{sec:query_decomposition}
Let $Q$ be a hot pattern that {\logo} decides to redistribute and let $D_Q$ be the data accessed by this pattern. Our goal is to redistribute (partition) $D_Q$ among all workers such that $D_Q$ can be evaluated without communication. 
Unlike previous work that performs static MinCut-based partitioning \cite{metis},
we eliminate the edge cuts by replicating edges that cross partition boundaries. 
Since the partitioning is an NP-complete problem, we introduce a heuristic for partitioning $D_Q$ with two objectives in mind: \myNum{i} the redistribution of $D_Q$ should benefit $Q$ as well as other pattens. \myNum{ii} Because replication is necessary for eliminating communication, redistributing $D_Q$ should result in minimal replication.

\begin{figure}[t]
	\centering
	\includegraphics[width=0.7\linewidth]{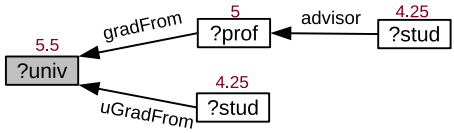}
	\caption{The query in Figure \ref{fig:score_example} transformed into a tree using Algorithm~\ref{algorithm:graph_walk}. 
	Numbers near vertices define their scores. The shaded vertex is the core.}
	\label{fig:query_decomposed}
\end{figure}

To address the first objective, we transform the patten $Q$ into a tree $T$ by breaking cycles and duplicating some vertices in the cycles. The reason is that cycles constrain the data grouped around the core to be also cyclic. For example, the query pattern in Figure \ref{fig:score_example} retrieves students who share the same alma mater with their advisors. Grouping the data around universities without removing the cycle is not useful for retrieving professors and their advisees who do not share the same university. Consequently, the pattern in Figure \ref{fig:score_example} can be transformed into a tree by breaking the cycle and duplicating the $?stud$ vertex as shown in Figure \ref{fig:query_decomposed}. We refer to the result of the transformation as {\em redistribution tree}.

Our goal is to construct the redistribution tree that minimizes the expected amount of replication. In Section~\ref{sec:vertexScore}, we explained why starting from the vertex with the highest score has the potential to minimize replication. Intuitively, the same idea applies recursively to each level of the redistribution i.e., every child node in the tree has a lower score than its parent. Obviously, this cannot be always achieved; for example in a path pattern where a lower score vertex comes between two high score vertices. Therefore, we use a greedy algorithm for transforming a hot pattern $Q$ into a redistribution tree $T$. Specifically, using the scoring function discussed in the previous section, we
first transform $Q$ into a vertex weighted, undirected graph $G$, where each node has a score and the directions of edges in $Q$ are disregarded.
The vertex with the highest score is selected as the core vertex. Then, $G$ is transformed into the redistribution tree using Algorithm~\ref{algorithm:graph_walk}.

Algorithm~\ref{algorithm:graph_walk} is a modified version of the Breadth-First-Search (BFS) algorithm, which has the following differences:
\myNum{i} unlike BFS trees which span all vertices in the graph, our tree span all edges in the graph. Each of the edges in the query graph should appear exactly once in the tree while vertices may be duplicated. \myNum{ii} During traversal, vertices with high scores are identified and explored first (using a priority queue). Since our traversal needs to span all edges, elements in the priority queue are stored as edges of the form ($parent$, $vertex$, $predicate$). These elements are ordered based on the vertex score first then on the edge label (predicate). Since the exploration does not follow the traditional BFS ordering, we maintain a pointer to the parent so edges can be inserted properly in the tree. As an example, consider the query in Figure \ref{fig:score_example}. Having the highest score, \emph{?univ}
is chosen as core, and the query is transformed into the tree shown in Figure~\ref{fig:query_decomposed}. Note that the nodes have weights (scores) and the directions of edges have been moved back.

\subsection{Incremental Redistribution}
\label{sec:incremental_redistribution}

Incremental ReDistribution (IRD) aims at redistributing data accessed by hot patterns among all workers in a way that eliminates communication while achieving high parallelism. Given a redistribution tree, {\logo} distributes the data along paths from the root to leaves using depth first traversal. The algorithm has two phases. First, it distributes triples containing the core vertex to workers using hash function $\mathcal{H}(\cdot)$.
Let $t$ be such a triple and let $t.core$
be its core vertex (the core can be either the subject or the object of $t$).
Let $w_1, \ldots, w_N$ be the workers. $t$ will be hash-distributed to worker $w_j$,
where $j = \mathcal{H}(t.core) \mod N$. 
Note that if $t.core$ is a subject, $t$ will not be replicated by IRD because of the initial subject-based hash partitioning.

\begin{table}\scriptsize
	\caption{Triples from Figure~\ref{fig:rdf_graph} matching patterns in Figure~\ref{fig:query_decomposed}.}
	\centering{
		\begin{tabular}{lc|l|l|lcll}
			\hline
			\multicolumn{4}{c|}{Worker 1} & \multicolumn{4}{c}{Worker 2} \\ \hline
			$t_1$& \multicolumn{3}{l|}{\RDF{Lisa}{uGradFrom}{MIT}} &      $t_3$ & \multicolumn{3}{l}{\RDF{Bill}{uGradFrom}{CMU}} \\
			& \multicolumn{3}{l|}{} &      $t_4$ & \multicolumn{3}{l}{\RDF{James}{uGradFrom}{CMU}} \\
			& \multicolumn{3}{l|}{} &       $t_5$& \multicolumn{3}{l}{\RDF{John}{uGradFrom}{CMU}} \\ \hline
			$t_2$& \multicolumn{3}{l|}{\RDF{James}{gradFrom}{MIT}} &       $t_6$& \multicolumn{3}{l}{\RDF{Bill}{gradFrom}{CMU}} \\ \hline
			$t_7$& \multicolumn{3}{l|}{\RDF{Lisa}{advisor}{James}} &       $t_8$& \multicolumn{3}{l}{\RDF{Fred}{advisor}{Bill}} \\
			& \multicolumn{3}{l|}{} &       $t_9$ & \multicolumn{3}{l}{\RDF{John}{advisor}{Bill}} \\
			& \multicolumn{3}{l|}{} &       $t_{10}$ & \multicolumn{3}{l}{\RDF{Lisa}{advisor}{Bill}} \\ \hline
		\end{tabular}
	}
	\label{table:exampleTriples}
\end{table}

In Figure~\ref{fig:query_decomposed}, consider the first-hop triple patterns \RDF{?prof}{uGradFrom}{?univ} and \RDF{?stud}{gradFrom}{?univ}. 
The core $?univ$ determines the placement of $t_1$-$t_6$ (see Table~\ref{table:exampleTriples}). Assuming two workers, $t_1$ and $t_2$ are hash-distributed to 
$w_1$ (because of MIT), whereas $t_3$-$t_6$ are hash-distributed to $w_2$ (because of CMU). 
The objects of triples $t_1$-$t_5$ 
are called their
\emph{source} columns.

\begin{mydef}[Source column] The source column of a triple 
is the column (subject or object) that determines its placement.
\end{mydef}

The second phase of the IRD places triples of the remaining levels of the tree in the workers that contain their parent triples,
through a series of distributed semi-joins. The column at the opposite end of the source column of the previous step 
becomes the \emph{propagating} column; in our previous example, the propagating column is the subject (i.e., $?prof$).

\begin{mydef} [Propagating column] The propagating column of a triple 
is its object (resp. subject) if the source column of the triple is its subject (resp. object).
\end{mydef}

At the second level of the redistribution tree in Figure~\ref{fig:query_decomposed}, the only subquery pattern is \RDF{?stud}{advisor}{?prof}. The propagating column $?prof$ from
the previous level becomes the source column for the current pattern. Triples $t_{7\ldots 10}$ in Table~\ref{table:exampleTriples} match
the sub-query and are joined with triples $t_{1\ldots 6}$. Accordingly, $t_7$ is placed in worker
$w_1$, whereas $t_7$, $t_9$ and $t_{10}$ are sent to $w_2$. 

\begin{figure*}
	\centering
	\includegraphics[width=2.1\columnwidth]{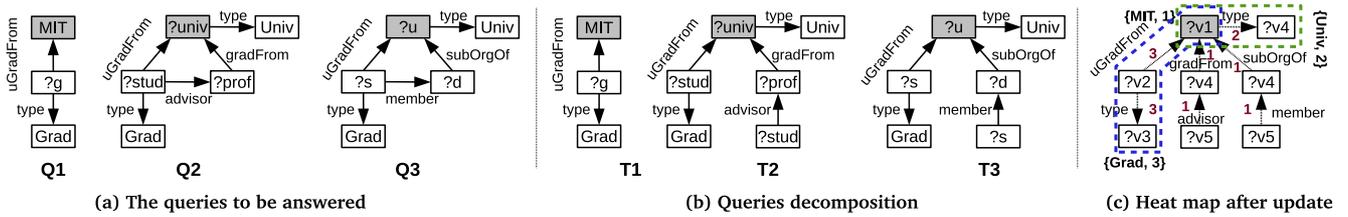}
	\caption{Updating the heat map. Selected areas indicate hot patterns.}
	\label{fig:heat_map}
\end{figure*}

\input{phd.tex}

The IRD process is formally described in Algorithm \ref{algorithm:phd}. For brevity, we describe the algorithm on a path input since we follow depth first traversal. The algorithm runs in parallel on all workers. Lines 1-5 hash distribute triples that contain the core vertex $\mathcal{C}$, if necessary.%
\footnote{\mbox{Recall if a core vertex is a subject, we do not redistribute.}}
Then, triples of the remaining levels are localized (replicated) in the workers that contain their parent. 
Replication is avoided for each triple which is already in the worker. 
This is carried out through a series of DSJ (lines 6-10). We maintain candidate triples in each level rather than final join results. Managing replicas in raw triple format allows us to utilize the RDF indices when answering queries using \mbox{replicated data.}

\vspace{-5pt}
\subsection{Queryload Monitoring}
\label{sec:monitoring}

\begin{figure*}
	\centering
	\includegraphics[width=1.7\columnwidth]{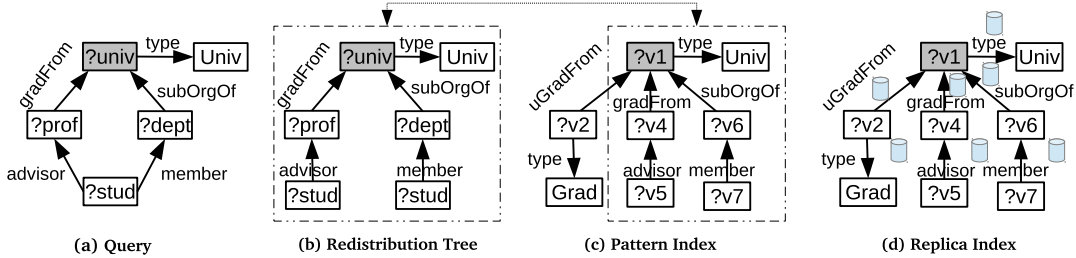}
	\caption{A query and the pattern index that allows execution without communication.}
	\label{fig:pi}
\end{figure*}

To effectively monitor workloads, systems face the following challenges: \myNum{i} the same query pattern may occur with different constants, subquery orderings, and variable names. Therefore, queries in the workload need to be deterministically transformed into a representation that unifies similar queries. 
\myNum{ii} This representation needs to be updated incrementally with minimal overhead. Finally, \myNum{iii} monitoring should be done at the level of patterns not whole queries. This allows the system to identify common hot patterns among \mbox{queries.}

\stitle{Heat map.}
We introduce a hierarchical heat map representation to monitor workloads. The heat map is maintained by the redistribution controller. Each query $Q$ is first decomposed into a redistribution tree using Algorithm~\ref{algorithm:graph_walk} (i.e., the procedure described in Section \ref{sec:query_decomposition}). The result is a tree $T$ with the core vertex as root. To detect overlap among queries, we transform $T$ to a tree template $\mathcal{T}$ in which all the constants are replaced with variables. To avoid loosing information about constant bindings in the workload, we store the constants values and their frequencies as meta-data in the template vertices. After that, $\mathcal{T}$ is inserted in the heat map which is a 
prefix-tree like structure that includes and combines the tree templates of all queries. Insertion proceeds by traversing the heat map from the root and matching edges in $\mathcal{T}$. If the edge does not exist, we insert a new edge in the heat map and set the edge count to 1; otherwise, we increment the edge count. Furthermore, we update the meta-data of vertices in the heat map with the meta-data in $\mathcal{T}$'s vertices. 
For example, consider queries $Q_1$, $Q_2$ and $Q_3$ and their decompositions $T_1$, $T_2$ and $T_3$, respectively in Figure \ref{fig:heat_map}(a) and (b). Assume that each of the queries is executed once. The state of the heat map after executing these queries is shown in Figure \ref{fig:heat_map}(c). Every inserted edge updates the edge count and the vertex meta-data in the heat map. For example, edge \RDF{$?v_2$}{uGradFrom}{$?v_1$} has edge count 3 because it appears in all $\mathcal{T}$'s. Furthermore, \{${MIT, 1}$\} is added to the meta-data of $v_1$. 

\stitle{Hot pattern detection.} The redistribution controller monitors queries by updating the heat map. As more queries are executed, the controller identifies hot patterns from the heat map. Currently, we use a hardwired
frequency threshold%
\footnote{Auto-tuning the frequency threshold is a subject of our future work.} for identifying hot patterns. Once a hot pattern is detected, the redistribution controller triggers the IRD process for that pattern. Remember that patterns in the heat map are templates in which all vertices are variables. To avoid excessive replication, some variables are replaced by dominating constants stored in the heat map. For example, assume the selected part of the heat map in Figure \ref{fig:heat_map}(c) is identified as hot. We replace vertex $?v_3$ with the constant Grad because it is the dominant value. On the other hand, $?v_1$ is not replaced by MIT because 
MIT does not dominate other values in query instances that include the hot pattern.
We use Boyer-Moore majority vote algorithm \cite{moore} for deciding the dominating constant.

\vspace{-5pt}
\subsection{Pattern and Replica Index}
\label{sec:rep_mgmt}

\stitle{Pattern index.}
The pattern index is created and maintained by the replication controller at the master. It has the same structure as the heat map, but it only stores redistributed patterns. For example, Figure \ref{fig:pi}(c), shows the pattern index state after redistributing all patterns in the heat map (Figure \ref{fig:heat_map}(c)). The pattern index is used by the query planner to check if a query can be executed without communication. When a new query $Q$ is posed, the planner transforms $Q$ into a tree $T$. If the root of $T$ is also a root in the the pattern index and all of $T$'s edges exist in the pattern index, then $Q$ can be answered in parallel mode; otherwise, $Q$ is answered in a distributed fashion.
For example, the query in Figure \ref{fig:pi}(a) can be answered in parallel because its redistribution tree (Figure \ref{fig:pi}(b)) is contained in the pattern index. Edges in the pattern index are time-stamped at every access to facilitate our eviction policy.

\stitle{Replica index.}
The replica index at each worker is identical to the pattern index at the master and is also updated by the IRD process. 
However, each edge in the replica index is associated with a storage module similar to the one that stores the original data. Each module stores only the replicated  data of the specified triple pattern. In other words, we do not add the replicated data to the main indices nor keep all replicated data in a single index. There are four reasons for this segregation. \myNum{i} As more patterns are redistributed, updating a single index becomes a bottleneck. \myNum{ii} Because of replication, using one index mandates filtering duplicate results.
\myNum{iii} If data is coupled in a single index, intermediate join results will be larger, which will affect performance. Finally, \myNum{iv} this hierarchical representation allows us to evict any part of the replicated data quickly without affecting the overall system performance.
Notice that we do not replicate data associated with triple patterns whose subjects are core vertices. Such data are accessed from the main index directly because of the initial subject-based hash partitioning. Figure \ref{fig:pi}(d) shows the replica index that has the same structure as the pattern index in Figure \ref{fig:pi}(c). The storage module associated with \RDF{?v7}{member}{?v6} stores replicated triples that match the triple pattern. Moreover, these triples qualify for the join with the triple patern of the parent edge.

\stitle{Conflicting Replication and Eviction.}
Conflicts arise when a subquery appears at different levels in the pattern index. This may cause some triples to be replicated by the hot patterns that include them. In terms of correctness, this is not a problem for {\logo} as conflicting triples (if any) are stored separately using different storage modules. This approach avoids the burden of any housekeeping and management of duplicates at the cost of memory consumption. Nevertheless, {\logo} employs an \emph{LRU} eviction policy that keeps the system within a given replication budget at each worker. 


%% file: graph_walk.tex
\begin{algorithm}[t]
	\scriptsize
	\KwIn{$G = \{V,E\}$; a vertex-weighted, undirected graph, the core vertex $v'$}
	\KwResult{The redistribution tree $T$}
	\SetKw{Let}{Let}
	\SetKwFunction{push}{$edges$.push}
	\SetKwFunction{pop}{$edges$.pop}
	\SetKwFunction{add}{$T$.add}
	\SetKwFunction{remove}{$verts$.remove}
	\Let $edges$ be a priority queue of pending edges\\
	\Let $verts$ be a set of pending vertices\\
	\Let $core\_edges$ be all incident edges to $v'$\\
	$visited[v']$ = true\;
	$T.root$=$v'$\;
	\ForEach{$e$ in $core\_edges$}{
		\push{$v'$, $e.nbr$, $e.pred$}\;
		$verts$.insert($e.nbr$)\;
		\add{$v', e.pred, e.nbr$}\;
	}
	\While{$edges$ notEmpty}{
		$(parent, vertex, predicate) \gets$ \pop{}\;
		$visited[vertex]$ = true\;
		\remove{$vertex$}\;
		\ForEach{$e$ in $vertex.edges$}{
			\If{$e.nbr$ NOT visited}{
				\If{$e.nbr \notin verts$}{
					\push{$vertex, e.nbr, e.pred$}\;
					$verts$.insert($e.nbr$)\;
					\add{$vertex, e.pred, e.nbr$}\;
				}
				\Else{
					\add{$vertex, e.pred, duplicate(e.nbr)$}\;
				}
			}
		}
	}
	\caption{Pattern Transformation}
	\label{algorithm:graph_walk}
\end{algorithm}

%

%% file: phd.tex
\begin{algorithm}
 	\scriptsize
	\KwIn{$P = \{E\}$; a path of consecutive edges, $\mathcal{C}$ is the core vertex.}
	\KwResult{Data replicated along path $P$}
	\SetKwFunction{getTriples}{getTriplesOfSubQuery}
	\SetKwFunction{send}{sendToWorker}
	\SetKwFunction{dsj}{DSJ}
	\SetKwFunction{addData}{IndexCandidateTriples}
	\SetKwComment{cc}{// }{}
	
	\cc{hash-distributing the first (core-adjacent) edge}
	\If{$e_0$ is not replicated}{
		$coreData$ = \getTriples{$e_0$}\;
		\ForEach{$t$ in $coreData$}{
			$m$ = $B(\mathcal{C}) \mod N$; \cc{$N$ is the number of workers}
			\send{$t$, $m$}\;
		}	
	}
	\cc{then collocate triples from other levels}
	\ForEach{$i:$ $1 \rightarrow |E|$}{
		\If{$e_i$ is not replicated}{
			$candidTriples$ = \dsj{$e_0$, $e_i$}\;
			\addData{$candidTriples$}\;
		}
		$e_0$ = $e_i$;\
	}
	\caption{Incremental Redistribution}
	\label{algorithm:phd}
\end{algorithm}

%% file: sec_experimental.tex
\begin{table*}\scriptsize
\centering{
\caption{Datasets Statistics in millions (M)}
\resizebox{2\columnwidth}{!}{
\label{tab:datasets}
\begin{tabular}{ l r r r r r r r}
 \hline
    \multicolumn{1}{l}{\textbf{Dataset}} & \multicolumn{1}{c}{\textbf{Triples (M)}} &  \multicolumn{1}{c}{\textbf{\#S (M)}} & \multicolumn{1}{c}{\textbf{\#O (M)}} & \multicolumn{1}{c}{\textbf{\#S$\cap$O (M)}} & \multicolumn{1}{c}{\textbf{\#P}} & \multicolumn{1}{c}{\textbf{Indegree (Avg/StDev)}} & \multicolumn{1}{c}{\textbf{Outdegree (Avg/StDev)}}\\ \hline
    \textbf{LUBM-10240} & 1,366.71 & 222.21 & 165.29 & 51.00 & 18 & 16.54/26000.00 & 12.30/5.97\\
    \textbf{WatDiv}  & 109.23  & 5.21 & 17.93 & 4.72 & 86 & 22.49/960.44 & 42.20/89.25  \\ 
    \textbf{WatDiv-1B} & 1,092.16 & 52.12 & 179.09 & 46.95 & 86 & 23.69/2783.40 & 41.91/89.05\\ 
    \textbf{YAGO2} & 295.85 & 10.12 & 52.34 & 1.77 & 98 & 10.87/5925.90 & 56.20/71.96\\ 
    \textbf{Bio2RDF} & 4,644.44 & 552.08 & 1,075.58 & 491.73 & 1,714 & 8.64/21110.00 & 16.83/195.44\\  \hline
\end{tabular}
}
}
\end{table*}
%

\section{Experimental Evaluation}
\label{sec:experimental}

We evaluate {\logo} 
against existing systems. 
We also include a non-adaptive version of our system,
referred to as {\logo}-NA,  
which does not include the features described in Section \ref{sec:adaptivity}.
In Section \ref{sec:exp:setup} 
we provide the details of the data, the hardware setup, and the competitors to our approach. In Section \ref{subsec:startup}, we demonstrate the low startup and initial replication overhead of {\logo} compared to all other systems. Then, in Section \ref{subsec:query_perf}, we 
apply queries with different complexities 
on different datasets
to show that 
\myNum{i} 
{\logo} leverages the subject-based hash locality to 
achieve better or similar performance compared to other systems
and
\myNum{ii} 
the adaptivity feature of {\logo} renders it several orders of magnitude faster than other systems. 
In Section \ref{subsec:adapt}, we conduct a detailed study of the effect and cost of  {\logo}'s adaptivity feature. The results show that our system 
adapts incrementally to workload changes with minimal overhead without resorting to full data repartitioning. 
Finally, in Section \ref{subsec:scalable}, we study the data and machine \mbox{scalability of {\logo}.}

\subsection{Setup and Competitors}\label{sec:exp:setup}
\nsstitle{Datasets: }We conducted our experiments using real and synthetic datasets of variable sizes. Table \ref{tab:datasets} describes these datasets, where \#S, \#P, and \#O denote respectively the numbers of unique subjects, predicates, and objects.
We use the synthetic LUBM\footnote{http://swat.cse.lehigh.edu/projects/lubm/} data generator to create a dataset of 10,240 universities consisting of 1.36 billion triples. WatDiv\footnote{http://db.uwaterloo.ca/watdiv/} is a recent benchmark that provides a wide spectrum of queries with varying structural characteristics and selectivity classes. We mainly used two versions of this dataset: WatDiv (109 million) and WatDiv-1B (1 billion) triples. LUBM and its template queries are usually used by most distributed RDF engines \cite{triad,shape,h2rdf2,trinity.rdf} for testing their query evaluation performance. However, LUBM queries are intended for semantic inferencing and their complexities lie in semantics not structure. Therefore, we also use WatDiv dataset which provides a wide range of query complexities and selectivity classes. As both LUBM and WatDiv are synthetic, we also use two real datasets. 
YAGO2\footnote{http://yago-knowledge.org/} is a real dataset derived from Wikipedia, WordNet and GeoNames containing 300 million triples. 
Bio2RDF dataset provides linked data for life sciences using semantic web technologies. We use Bio2RDF\footnote{http://download.bio2rdf.org/release/2/} release 2, which contains 4.64 billion triples connecting 24 different \mbox{biological datasets.}

\nsstitle{Hardware Setup: }
We implemented {\logo} in C++ and used a Message Passing Interface library (MPICH2) for synchronization and communication. Unless otherwise stated, we deploy {\logo} and its competitors on a cluster of 12 machines each with 148GB RAM and two 2.1GHz AMD Opteron 6172 CPUs (12 cores each). The machines run 64-bit 3.2.0-38 Linux Kernel and are connected by a 10Gbps Ethernet switch. 

\nsstitle{Competitors: }
We compare our framework against two recent in-memory RDF systems, Trinity.RDF \cite{trinity.rdf} and TriAD \cite{triad}. 
To the best of our knowledge, these systems provide the fastest query response times. However, they were not available to us for comparison; the only way to compare against them is to use the reported runtimes in the corresponding papers \cite{trinity.rdf,triad}. Note that our testbed is slightly inferior to those used in \cite{trinity.rdf,triad}. In particular, Trinity.RDF uses 40Gbps InfiniBand interconnect, which is theoretically 4X faster than ours. TriAD uses faster processors with a larger number of cores interconnected with a slower interconnect (1Gbps Ethernet). Nonetheless, because of its sophisticated partitioning scheme and join-ahead pruning, TriAD communicates small amounts of data during query evaluation (tens of Megabytes). 
Therefore, using a faster interconnect is not going to affect its performance significantly on the datasets they used.

We also compare with two Hadoop-based systems that employ lightweight partitioning: SHARD \cite{Rohloff10:shard} and H2RDF+ \cite{h2rdf2}. Furthermore, we compare to SHAPE
\footnote{SHAPE showed better replication and query performance than H-RDF-3X \cite{Huang11:scalable}. Hence, we only compare to SHAPE.}
\cite{shape}, a system that relies on static replication and uses RDF-3X as underlying data store. We limit our comparison to distributed systems only, because they 
outperform state-of-the-art centralized RDF systems. 

\vspace{-10pt}
\subsection{Startup Time and Initial Replication}
\label{subsec:startup}

Our first experiment measures the time it takes all systems for preparing the data prior to answering queries. We exclude the string-to-id mapping time for all systems. For TriAD, we show the time to partition the graph using METIS \cite{metis}. We used the same number of partitions reported in \cite{triad} for partitioning LUBM-10240 and WatDiv. The Bio2RDF and YAGO2 datasets are partitioned into 200K and 38K partitions, respectively.  As Table \ref{tab:preproc} shows, METIS is prohibitively expensive and does not scale for large RDF graphs. To apply METIS, we had to remove all triples connected to literals; otherwise, METIS takes several days to partition LUBM-10240, Bio2RDF and YAGO2 datasets. 

\begin{table}\scriptsize
\centering
\caption{Preprocessing time (minutes) }
\resizebox{\columnwidth}{!}{
\begin{tabular}{l r r r r}
 \hline
     & \multicolumn{1}{c}{\textbf{LUBM-10240}} & \multicolumn{1}{c}{\textbf{WatDiv}} & \multicolumn{1}{c}{\textbf{Bio2RDF}}& \multicolumn{1}{c}{\textbf{YAGO2}}\\ \hline
    \textbf{{\logo}} 	&\bfseries{14}		&\bfseries{1.2}		&\bfseries{115}		& \bfseries{4}   \\ 
    \textbf{METIS} 	& 523 			& 66 			& 4,532			& 105 \\ 
    \textbf{SHAPE} 	& 263 			& 79 			& $>$24h		& 251  \\ 
    \textbf{SHARD} 	& 72 			& 9 			& 143			& 17\\ 
    \textbf{H2RDF+} 	& 152 			& 9 			& 387			& 22\\ \hline
\end{tabular}
}
\label{tab:preproc}
\end{table}

We configured SHAPE with full level semantic hash partitioning and enabled the type optimization (see \cite{shape} for details). Furthermore, for fair comparison, SHAPE is configured to partition each dataset such that all its queries are processable without communication. For LUBM-10240, SHAPE incurs less preprocessing time compared to METIS-based systems. However, for WatDiv and YAGO2, SHAPE performs worse because of data imbalance, causing some of the RDF-3X engines to take more time in building the databases. Particularly, partitioning YAGO2 and WatDiv using 2-hop forward and 3-hop undirected, respectively, placed all the data in a single partition. The reason of this behavior is that all these datasets have uniform URI's and hence SHAPE could not fully utilize its semantic hash partitioning. SHAPE did not finish the partitioning phase of Bio2RDF and was terminated after 24 hours. 

SHARD and H2RDF+ employ lightweight partitioning, random and range-based, respectively. Therefore, they require less time compared to other systems. However, since they are Hadoop-based, they suffer from the overhead of storing the data first on Hadoop File System (HDFS) before building their data stores.

{\logo} uses lightweight hash partitioning and avoids the upfront cost of sophisticated partitioning schemes. As Table \ref{tab:preproc} shows, {\logo} starts 4X up to two orders of magnitude faster than existing systems.

We only report the initial replication of SHAPE and TriAD,
since {\logo}, SHARD and H2RDF+ do not incur any initial replication (the replication caused by {\logo}'s adaptivity is evaluated in the next section). TriAD replicates all edges that cross partitions boundaries; producing a 1-hop undirected guarantee. Therefore, we consider the edge-cut reported by METIS to be the amount of replication in TriAD. Table~\ref{tab:initial_replication} shows the replication ratio as a percentage of the original data size. 
For LUBM-10240, TriAD results in the least replication as LUBM is uniformly structured around universities. 
With full level semantic hash partitioning and type optimization, SHAPE incurs almost double the replication of TriAd. For WatDiv, METIS produces very bad partitioning because of the dense nature of the data. Consequently, TriAD results in excessive replication because of the high edge-cut. Note that the highest radius in all WatDiv query templates is 3 (undirected); and partitioning the whole data blindly using $k$-hop guarantee as in H-RDF-3X \cite{Huang11:scalable} will result in excessive replication which grows exponentially as $k$ increases. The same thing applies to Bio2RDF and YAGO2 datasets. SHAPE places the data on a single partition because of the URI's uniformity of WatDiv and YAGO2. Therefore, it incurs no replication but performs as good as a single machine RDF-3X store. 

\begin{table}\scriptsize
\centering
\caption{Initial replication}
\label{tab:initial_replication}
\resizebox{\columnwidth}{!}{
\begin{tabular}{lrrrr}
\hline
			& \textbf{LUBM-10240} 		& \textbf{WatDiv} 	& \textbf{Bio2RDF} & \textbf{YAGO2}	\\ \hline
\textbf{SHAPE}    	& 42.9\%           		& (1 worker) 0\% 	& NA		& (1 worker) 0	\\ 
\textbf{TriAD}    	& 23.6\%              		& 82.9\%   		& 30.0\%		 & 40.0\%			\\ \hline
\end{tabular}
}
\end{table}

\vspace{-10pt}
\subsection{Query Performance}
\label{subsec:query_perf}
In this section, we compare {\logo} performance on individual queries against state-of-the-art distributed RDF systems using multiple real and synthetic datasets. We demonstrate that even the  {\logo}-NA version of our system (which does not include the adaptivity feature) 
is competitive in performance to systems that employ sophisticated partitioning techniques.  This shows that the subject-based hash partitioning and the distributed evaluation techniques proposed in Section \ref{sec:distributed_query} are very effective.
When {\logo} adapts, its performance becomes even better and our system consistently outperforms its competitors by a wide margin.

\begin{table}\scriptsize
\centering
\caption{Query runtimes for LUBM-10240 (ms)}
\label{my-label}
\resizebox{\columnwidth}{!}{%
\begin{tabular}{lrrrrrrr}
\hline
LUBM-10240           	& \textbf{L1} 		& \textbf{L2} 		& \textbf{L3} 		& \textbf{L4} 		& \textbf{L5} 		& \textbf{L6} 		& \textbf{L7} 		\\ \hline
\textbf{AdHash}   	& \textbf{317}		& \textbf{120} 		& \textbf{6}		& \textbf{1}  		& \textbf{1}   		& 4       		& \textbf{220}	\\
\textbf{AdHash-NA}  	& 2,743        		& 120        		& 320         		& \bfseries{1}  	& \textbf{1}    	& 40       		& 3,203        		\\
\textbf{SHAPE}       	& 25,319        	& 4,387        		& 25,360       		& 1,603        		& 1,574        		& 1,567        		& 15,026       		\\
\textbf{H2RDF+}      	& 285,430       	& 71,720        	& 264,780      		& 24,120       		& 4,760        		& 22,910        	& 180,320      		\\
\textbf{SHARD}       	& 413,720       	& 187,310       	& aborted     		& 358,200      		& 116,620       	& 209,800       	& 469,340      		\\ \hline \hline
\textbf{TriAD-SG}    	& 2,146        		& 2,025         	& 1,647        		& \bfseries{1}  	& \textbf{1}    	& \textbf{1}    	& 16,863       		\\
\textbf{Trinity.RDF} 	& 7,000         	& 3,500        		& 6,000        		& 4       		& 3       		& 10     		& 27,500       		\\
\hline
\end{tabular}}
\label{tab:lubm10240}
\end{table}

\stitle{LUBM dataset: } 
In the first experiment (Table \ref{tab:lubm10240}), we compare the performance of all systems using the LUBM-10240 dataset and queries L1-L7 defined in \cite{atre} and also 
used by Trinity.RDF and TriAD.%
\footnote{Recall from Section \ref{sec:exp:setup} that the numbers for Trinity.RDF \cite{trinity.rdf} and TriAD \cite{triad} are copied from their corresponding papers because these systems are not publicly available. Therefore, we only compare to them using the queries they used.}
For SHAPE to execute all these queries without communication, we use 2-hop forward semantic hash partitioning with the type optimization. Queries can be classified based on their structure and selectivities into simple and complex. L4 and L5 are simple selective star queries whereas L2 is a simple yet non-selective star query that generates large final results. L6 is considered as a simple query because it is highly selective. On the other hand, L1, L3 and L7 are complex queries that generate large intermediate results but return very small final results. 

SHARD and H2RDF+ suffer from the expensive overhead of MapReduce; hence, their performance is significantly worse than all other systems. On the other hand, SHAPE incurs minimal communication and performs better than SHARD and H2RDF+ due to the utilization of semantic hash partitioning. Because it uses MapReduce for dispatching queries to workers, it still suffers from the non-negligible overhead of MapReduce.

In-memory RDF engines, Trinity.RDF, TriAD-SG and AdHash, perform significantly better than systems based on MapReduce. 
Queries L4 and L5 are selective subject star queries that produce very small intermediate results. 
Therefore, in-memory systems can solve these queries efficiently. 
{\logo} exploits the initial hash distribution and solve these queries without communication, which explains why both versions of AdHash have the same performance. 
Similarly, L2 consists of a single subject-subject join; however, AdHash is faster than TriAD-SG and Trinity.RDF by more than an order of magnitude. 
Due to L2 low selectivity, the exploration of Trinity.RDF does not reduce the intermediate results size leading to an expensive centralized join by the master. 
TriAD, on the other hand, solves the query by two distributed index scans (one for each base subquery) followed by a distributed merge join. {\logo} performs better than TriAD-SG by avoiding unnecessary scans. In other words, utilizing its hash indexes and the right deep tree planning, {\logo} requires a single scan followed by hash lookups. 

TriAD's pruning technique eliminates the communication required for solving L6. Therefore, it significantly outperforms Trinity.RDF and {\logo}-NA. However, once {\logo} adapts, L6 is executed without communication resulting in a comparable performance to TriAD.

{\logo} outperforms Trinity.RDF and TriAD-SG for L1, L3 and L7. Even with simple hash partitioning, {\logo}-NA achieves better or comparable performance to both in-memory systems. Particularly, since these queries are cyclic, Trinity.RDF can not reduce the size of the intermediate results of these queries. All workers need to ship their intermediate results to the master to finalize the query evaluation in a centralized manner. Therefore, the master node is a potential bottleneck especially when the intermediate results are huge, like in L7. Even with the sophisticated partitioning and pruning technique of TriAD-SG, these queries still require inter-worker communication whereas {\logo} executes these queries in parallel without communication. For L3, {\logo}-NA is 5X to several orders of magnitude faster than TriAD-SG and Trinity.RDF. {\logo}-NA evaluates the join that leads to an empty intermediate results early causing {\logo}-NA to avoid useless joins. However, the first few joins cannot be eliminated during query planning time. On the other hand, {\logo} can detect queries with empty results during planning. As each worker makes its local parallel query plan, workers detects that the cardinality of the subquery in the replica index is zero and terminates.

\begin{table}\scriptsize
\centering
\caption{Query runtimes for WatDiv (ms)}
\label{tab:watdiv}
\resizebox{\columnwidth}{!}{%
\begin{tabular}{lrrrrr}
\hline
\textbf{WatDiv-100}        	& \textbf{Machines}	& \textbf{L1-L5} 	& \textbf{S1-S7} 	& \textbf{F1-F5} 	& \textbf{C1-C3} \\ \hline
\textbf{AdHash} 	& 5			& \textbf{2}           	& \textbf{1} 		& \textbf{10}  		& \textbf{12}    \\
\textbf{AdHash-NA} 	& 5			& 9			& 6			& 235			& 123            \\
\textbf{SHAPE}     	& 12			& 1,870           	& 1,824           	& 1,836           	& 2,723           \\
\textbf{H2RDF+}    	& 12			& 5,441           	& 8,679           	& 18,457          	& 65,786          \\ \hline \hline
\textbf{TriAD}  	& 5			& \textbf{2}           	& 3              	& 29             	& 270            \\
\hline
\end{tabular}}
\end{table}

\stitle{WatDiv dataset: }The WatDiv benchmark defines 20 query templates\footnote{\url{http://db.uwaterloo.ca/watdiv/basic-testing.shtml}} classified into four categories: linear (L), star (S), snowflake (F) and complex queries (C). Similar to TriAD, we generated 20 queries using the WatDiv query generator for each query category C, F, L and S. We deployed {\logo} on five machines to match the setting of TriAD in \cite{triad}. Table \ref{tab:watdiv} shows the performance of {\logo} compared to other systems. For each complexity family, we calculate the geometric mean of each system. H2RDF+ performs worse than all other systems due to the overhead of MapReduce. 
SHAPE, under 2-hop forward partitioning, placed all the data in one machine; therefore, its performance is no better than a single-machine RDF-3X. {\logo} and TriAD, on the other hand, provide significantly better performance than MapReduce-based systems. TriAD benefits from its asynchronous message passing and performs better than {\logo}-NA in L, S, and F queries. For complex queries with large diameters {\logo}-NA performs better as a result of the locality awareness. When {\logo} adapts, it consistently performs better than all systems for all queries.

\begin{table}\scriptsize
\centering
\caption{Query runtimes for YAGO2 (ms)}
\label{tab:yago}
\resizebox{\columnwidth}{!}{%
\begin{tabular}{lrrrr}
\hline
YAGO2              	& \textbf{Y1} & \textbf{Y2} & \textbf{Y3} & \textbf{Y4} \\ \hline
\textbf{AdHash} 	&\textbf{2.5} &\textbf{19}  & \textbf{11} & \textbf{2}  \\
\textbf{AdHash-NA}    	& 19          & 46          & 570         & 77          \\
\textbf{SHAPE}     	& 1,824        & 665,514      & 1,823        & 1,871        \\
\textbf{H2RDF+}    	& 10,962       & 12,349       & 43,868       & 35,517       \\
\textbf{SHARD}     	& 238,861      & 238,861      & aborted     & aborted     \\ \hline
\end{tabular}
}
\end{table}

\stitle{YAGO dataset}
YAGO2 does not provide benchmark queries, therefore we created a set of representative 
test queries (Y1-Y4) defined in Appendix \ref{app:yago_queries}. We show in Table \ref{tab:yago} the performance of {\logo} against SHAPE, H2RDF+ and SHARD. {\logo}-NA continues to significantly outperform other systems for all queries. Furthermore, our adaptive version, {\logo}, is up to two orders of magnitude faster than all other systems.

\begin{table}\scriptsize
\centering
\caption{Query runtimes for Bio2RDF (ms)}
\label{tab:bio}
\resizebox{\columnwidth}{!}{%
\begin{tabular}{lrrrrr}
\hline
\textbf{Bio2RDF}	& \textbf{B1} 		& \textbf{B2} 		& \textbf{B3}		& \textbf{B4}		& \textbf{B5} 		\\ \hline
\textbf{AdHash} 	&\textbf{4}		&\textbf{2}		&\textbf{2}		&\textbf{2}		&\textbf{1}	 	\\
\textbf{AdHash-NA}    	&19			&16			&36			&187			&1			\\
\textbf{H2RDF+}    	&5,580			&12,710			&322,300		&7,960			&4,280			\\
\textbf{SHARD}     	&239,350		&309,440		&512,850		&787,100		&112,280		\\ \hline
\end{tabular}}
\end{table}

\stitle{Bio2RDF dataset:} Similar to YAGO2 dataset, the Bio2RDF dataset does not have benchmark queries; therefore, we defined five queries (B1-B5) that have different structures and complexities. B1 requires object-object join which contradicts our initial data distribution. Queries B2, B3 are star queries with different number of triple patterns that require subject-object and/or subject-subject joins. B5 is a simple star query with only one triple pattern while B4 is a complex query with 2-hops radius.
We could not run SHAPE as it failed to preprocess the data using 2-hop forward partitioning within reasonable time. Similar to their behavior in LUBM-10240 and WatDiv datasets, H2RDF+ and SHARD still are worse than {\logo} due to the MapReduce overhead. Overall, {\logo} outperforms all other systems by orders of magnitude. 

%

\begin{figure}
\centering%
  \subfigure[Execution Time]{
  	\label{fig:lubm_locality_time}
  	\includegraphics[width=0.63\columnwidth]{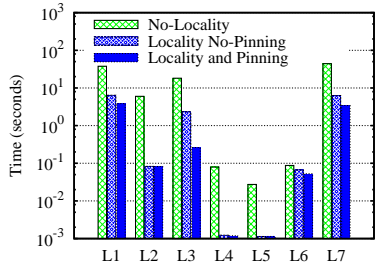}
  }
  \subfigure[Communication Cost]{
  	\label{fig:lubm_locality_comm}
  	\includegraphics[width=0.63\columnwidth]{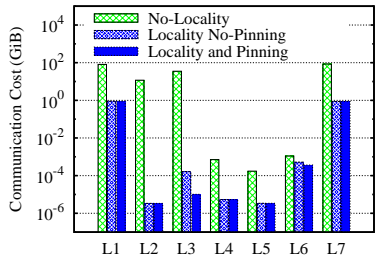}
  }%
  \caption{Impact of locality awareness on LUBM-10240.}
\end{figure}

\vspace{-2pt}
\subsubsection{Impact of Locality Awareness} 

In this experiment, we show the effect of locality aware planning on the distributed query evaluation of {\logo}-NA (non-adaptive). We define three configurations of {\logo}: \myNum{i} We disable the $pinned\_subject$ optimization and hash locality awareness. \myNum{ii} We disable the $pinned\_subject$ optimization while maintaining the hash locality awareness; in other words, workers can still know the locality of subject vertices but joins on the pinned subjects are synchronized. Finally, \myNum{iii} we enable all optimizations. We run the LUBM (L1-L7) queries on the LUBM-10240 dataset on all configurations of {\logo}-NA. The query response times and the communication costs are shown in Figures \ref{fig:lubm_locality_time} and \ref{fig:lubm_locality_comm}, respectively. Disabling hash locality resulted in excessive communication which drastically affected the query response times. Enabling the hash locality affected all queries except L6 because of its high selectivity. The performance gain for other queries ranges from 6X up to 2 orders of magnitude. In the third configuration, the pinned subject optimization does not affect the amount of communication because of the hash locality awareness. In other words, since the joining subject is local, {\logo} does not communicate intermediate results. However, performance is affected by the synchronization overhead. Queries like L2, L4 and L5 are not affected by this optimization because they are star queries joining on the subject. On the other hand, all queries that require communication are affected. The performance gain ranges from 26\% in case of L6 to more than 90\% for L3. The same behavior is also noticed in the WatDiv-1B dataset.

\begin{figure*}\centering
  \subfigure[Execution time]{
  	\label{exp:time_sens}
  	\includegraphics[width=0.63\columnwidth]{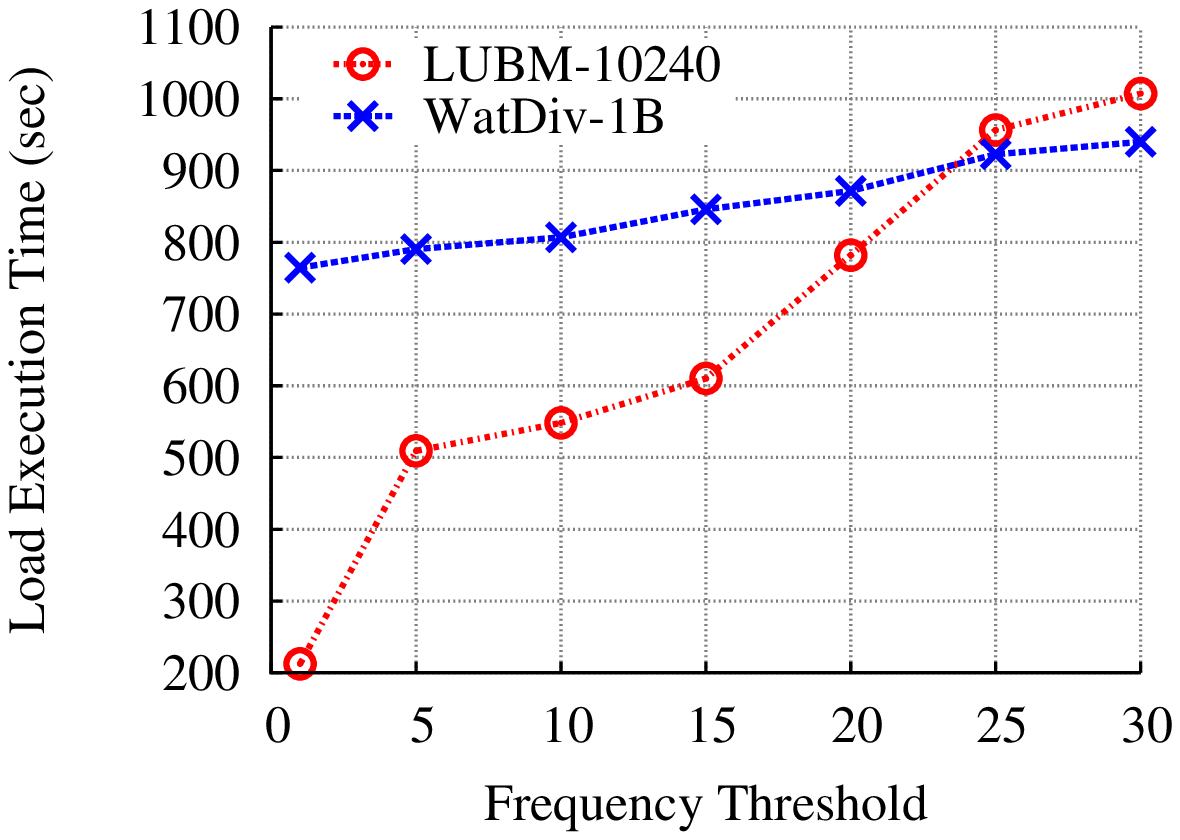}
  }\hfill
  \subfigure[Communication cost]{
  	\label{exp:comm_sens}
  	\includegraphics[width=0.63\columnwidth]{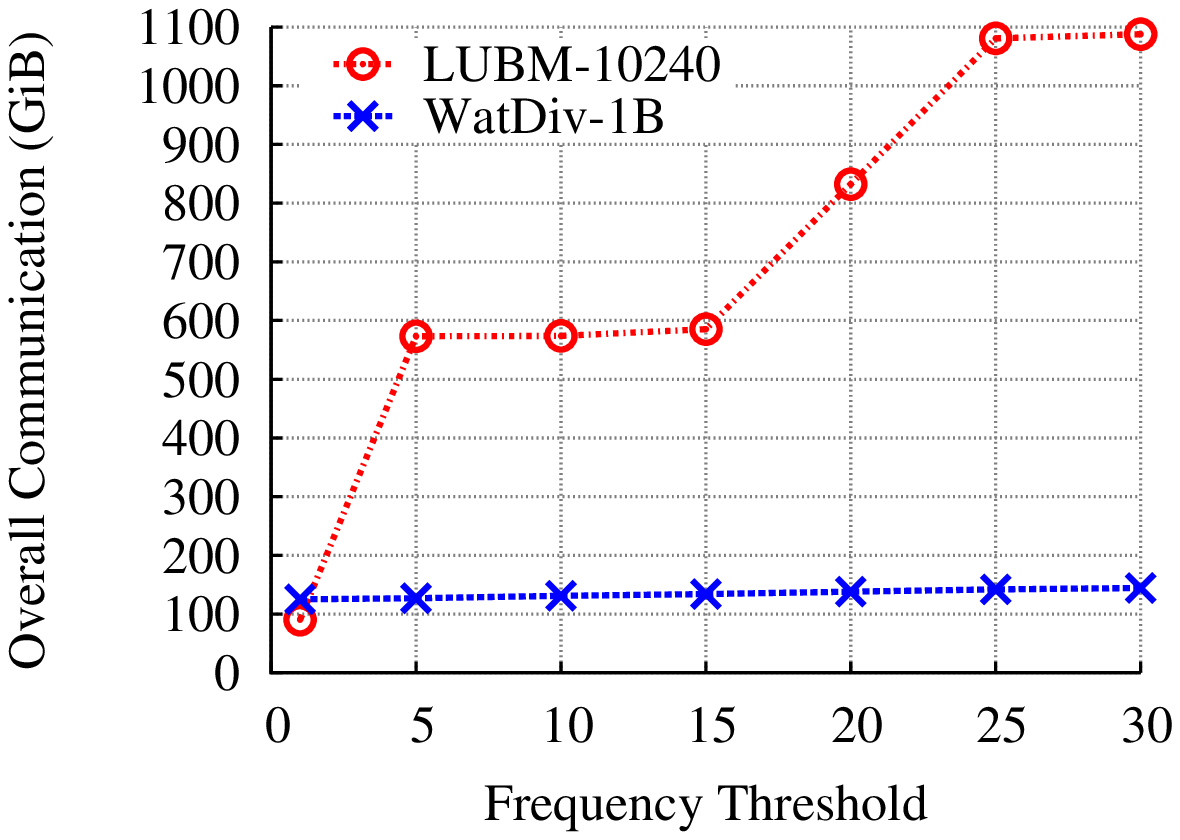}
  }\hfill
  \subfigure[Replication cost]{
  	\label{exp:rep_sens}
  	\includegraphics[width=0.63\columnwidth]{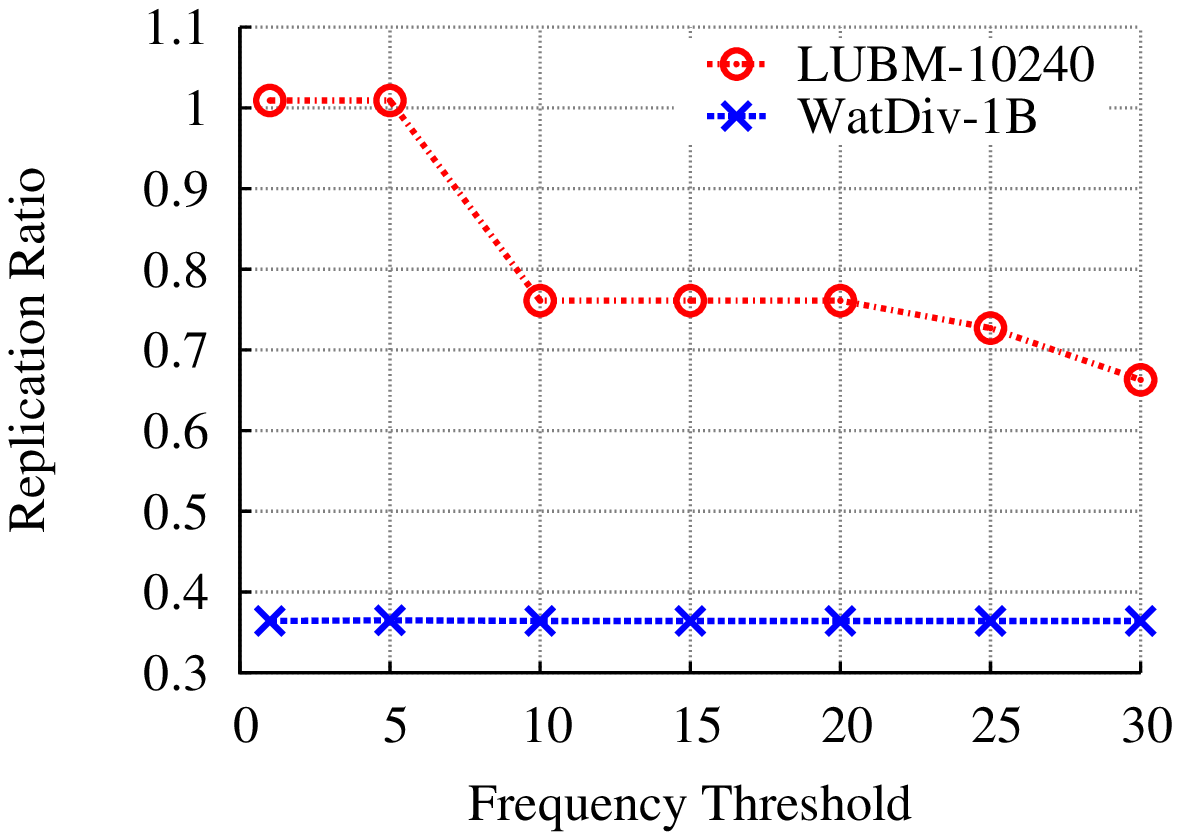}
  }
  \caption{Frequency threshold sensitivity analysis.}
\end{figure*}
\vspace{-15pt}
\subsection{Workload Adaptivity by {\logo}}
\label{subsec:adapt}

In this section, we thoroughly evaluate {\logo}'s adaptivity. For this purpose, we define different workloads on two billion scale datasets with different characteristics, namely, WatDiv-1B and LUBM-10240.\\
\nsstitle{WatDiv-1B workload: }The WatDiv benchmark defines 20 query templates\footnote{\url{http://db.uwaterloo.ca/watdiv/basic-testing.shtml}} classified into four categories: linear (L), star (S), snowflake (F) and complex queries (C). We used the benchmark query generator to create a 5K-queries workload from each category, resulting in a total of 20K queries. We also generated a random workload of 20K queries from all query templates. 

\nsstitle{LUBM-10240 workload: }As {\logo} and the other systems do not support inferencing, we used all 14 queries in the LUBM benchmark without reasoning\footnote{Only query patterns are used. Classes and properties are fixed so queries return large intermediate results.}. All queries are listed in Appendix \ref{app:lubm_queries}. 
From these queries, we generated 10K queries that have different constants. 
Then, we randomly selected 20K queries from the 10K queries. 
This workload covers a wide spectrum of query complexities including simple selective queries, star queries as well as queries with complex structures and low selectivities.
For details, refer to Appendix \ref{app:lubm_wload}.

\subsubsection{Frequency Threshold Sensitivity Analysis}

The frequency threshold controls the triggering of the Incremental ReDistribution (IRD) process. Consequently, it influences the execution time and the amount of communication and replication in the system. In this experiment, we conduct an empirical sensitivity analysis to select the frequency threshold value based on the two aforementioned query workloads. We execute each of the workloads while varying the frequency threshold values from 1 to 30. Note that our frequency monitoring is not on a query-by-query basis as our heat map monitors the frequency of the subquery pattern in a hierarchical manner (see Section \ref{sec:monitoring}). The workload execution times, the communication costs and the resulting replication ratios are shown in Figures~\ref{exp:time_sens}, \ref{exp:comm_sens} and \ref{exp:rep_sens}, respectively. 

We observe that LUBM-10240 is very sensitive to slight changes in the frequency threshold because of the complexity of its queries. As the frequency threshold increases, the redistribution of hot patterns is delayed causing more queries to be executed with communication. 
Consequently, the amount of communication and synchronization overhead in the system increases, affecting the overall execution time, while the replication ratio is low because fewer patterns are redistributed. 

On the other hand, WatDiv-1B is not as sensitive to this range of frequency threshold because most of its queries are solved in subseconds using our locality-aware distributed semi-join algorithm; and do not incur excessive communication. 
Nevertheless, as the frequency threshold increases, the synchronization overhead affects the overall execution time. 
Furthermore, due to our fine-grained query monitoring, {\logo} captures the commonalities between the WatDiv-1B query templates for frequency thresholds 5 to 30. Hence, for all these thresholds the replication ratio remains almost the same. The difference is that the system will converge faster for lower threshold values, reducing the overall execution time. 
In all subsequent experiments, we use a frequency threshold of 10 as it resulted in a good balance between time and replication. 
We plan to study the auto-tuning of this parameter in the future.

\subsubsection{Workload Execution Cost}

\begin{figure}
\centering
  \subfigure[Execution time]{
  	\label{fig:watdiv_adapting}
  	\includegraphics[width=0.63\columnwidth]{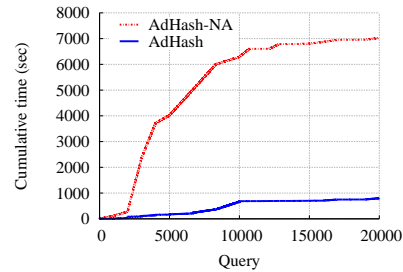}
  }
  \subfigure[Communication cost]{
  	\label{fig:watdiv_comm}
  	\includegraphics[width=0.63\columnwidth]{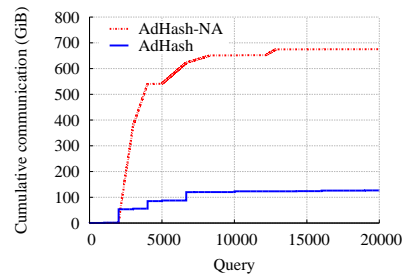}
  }
  \caption{{\logo} adapting to workload (WatDiv-1B).}
\end{figure}

To simulate a change in the workload, queries of the same WatDiv-1B template are run consecutively while enforcing a replication threshold of 20\% per worker. Figure \ref{fig:watdiv_adapting}  shows the cumulative time as the execution progresses with and without the adaptivity feature. After every sequence of 5K query executions, the type of queries changes. Without adaptivity (i.e.,  {\logo}-NA), the cumulative time increases sharply as long as complex queries are executed (e.g., from query 2K to query 10K). On the other hand, {\logo} adapts to the change in workload with little overhead causing the cumulative time to drop significantly by almost 6 times.

\begin{figure}
\centering
  \subfigure[Execution time]{
  	\label{fig:lubm_adapting}
  	\includegraphics[width=0.63\columnwidth]{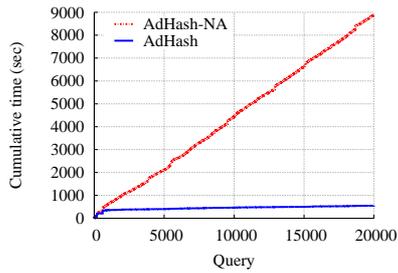}
  }
  \subfigure[Communication cost]{
  	\label{fig:lubm_comm}
  	\includegraphics[width=0.63\columnwidth]{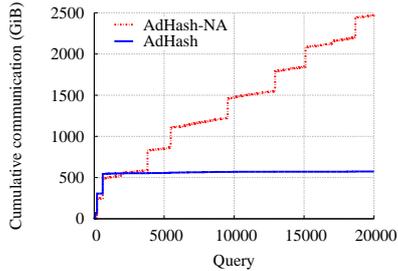}
  }
  \caption{{\logo} adapting to workload (LUBM-10240).}
\end{figure}

Figure \ref{fig:watdiv_comm} shows the cumulative communication costs of both {\logo} and {\logo}-NA. 
As we can see, the communication cost exhibits the same pattern as that of the runtime cost (Figure \ref{fig:watdiv_adapting}), which proves that communication and synchronization overheads are detrimental to the total query response time. The overall communication cost of {\logo} is more than 7X lower compared to that of {\logo}-NA. Once {\logo} starts adapting, most of future queries are solved with minimum or no communication. The same behavior is observed for the LUBM-10240 workload (see Figures \ref{fig:lubm_adapting} and \ref{fig:lubm_comm}).

\begin{figure}
\centering
  \subfigure[Execution time]{
  	\label{fig:partout_time}
  	\includegraphics[width=0.63\columnwidth]{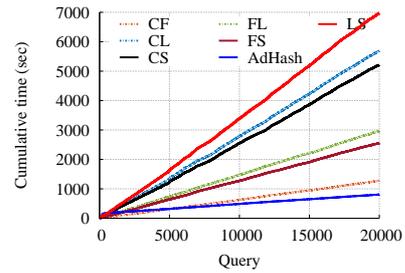}
  }
  \subfigure[Communication cost]{
  	\label{fig:partout_comm}
  	\includegraphics[width=0.63\columnwidth]{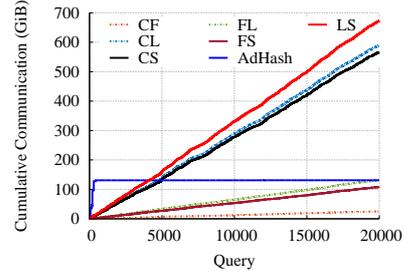}
  }
  \caption{Comparison with static representative workload-based partitioning.}
\end{figure}

\nsstitle{Partitioning based on a representative workload: }We tried to partition these datasets based on a representative workload using Partout \cite{partout}. However, it could not partition the data using a large workload within a reasonable time ($<$24 hours). Consequently, in this experiment, we simulate the effect of assuming a representative workload when partitioning the data using {\logo}. To do so, we train {\logo} using different combinations of the different workload categories defined by WatDiv-1B (C, F, S, and L). Each combination is made of two categories; effectively producing six combinations, mainly CF, CL, CS, FL, FS, and LS. After training {\logo}, we test the system using a random workload selected from all query categories, which consists of 20K queries. This way, some of the queries in the test workload would run in parallel while others (not in the representative workload) would require communication. In Figures \ref{fig:partout_time} and \ref{fig:partout_comm}, we show the cumulative execution time and communication, respectively, for the test workloads (i.e. excluding the training time). In the same figures, we show the performance of {\logo} without training. Obviously, the performance of the test workload highly depends on the complexity of the queries used in the training phase. For example, the complex (C) and snowflake (F) queries are the most expensive queries in the benchmark. Therefore, when the system is trained using the CF training workload, it performs much better than when trained using the LS workload. On the other hand, allowing the system to adapt incrementally and dynamically (without training) resulted in better performance when compared to all cases. {\logo} incurs more communication at the beginning because of the IRD process. However, it then converges to almost constant communication. CF workload requires less communication because the L and S queries (not in the training workload) do not require excessive data exchange. Nonetheless, the CF execution time keeps increasing due to the existence of communication and synchronization overheads.

\subsubsection{Redistribution Tree Generation}
\label{subsec:trans}

In this experiment, we evaluate our query transformation heuristic (Section \ref{sec:vertexScore}) against other two alternative approaches. Recall that when transforming a hot query pattern into the redistribution tree, we select the vertex with the highest score to be the tree root. Then the query is traversed from high score vertices to lower score ones. Therefore, we compare our heuristic (referred to \emph{High-Low} hereafter) to two different heuristics: \myNum{i} the vertex with the least vertex score is selected as core; then the query pattern is traversed be exploring vertices with lower scores first. We refer to this heuristic as \emph{Low-High}. We also compare to \myNum{ii} another approach that uses a different vertex scoring function where the score of a vertex in the hot query pattern is its out-degree. The pattern is then traversed from high score vertices to lower score ones. We refer to this approach as \emph{QDegree}. Note that the latter approach aims at minimizing the replication in a greedy manner by fully exploiting the initial hash partitioning. Recall that data that binds to triple patterns whose subject is a core are not replicated. 

\begin{figure}
\centering
  \subfigure[Replication and Communication cost]{
  	\label{fig:lubm_heuristics_stat}
  	\includegraphics[width=0.63\columnwidth]{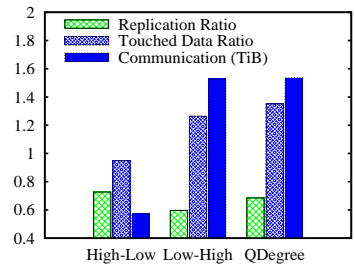}
  }
  \subfigure[Execution time]{
  	\label{fig:lubm_heuristics_time}
  	\includegraphics[width=0.63\columnwidth]{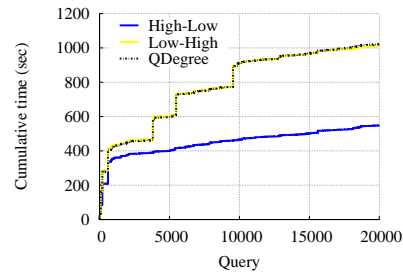}
  }
  \caption{Effect of hot pattern transformation.}
\end{figure}

We evaluated all these heuristics by running the LUBM-10240 workload. In Figure \ref{fig:lubm_heuristics_stat}, we show the resulting replication, the communication cost and the amount of data touched by the IRD process. The Low-High and the QDegree heuristics resulted in slightly less replication compared to the High-Low approach. The reason is that both heuristics benefit from the initial hash partitioning by selecting cores with larger number of outgoing edges. However, the amount of data touched by the redistribution process (i.e. data in the main and replica indices) in the Low-High and QDegree is significantly higher. This affects the adaptivity performance because the IRD process is carried out using a series of DSJ iterations. Furthermore, because the data touched by the process is actually used for evaluating parallel queries, the performance of parallel queries is eventually affected. 

Consequently, the cumulative workload execution time using the High-Low heuristic is 1.9X faster than the other heuristics as shown in Figure \ref{fig:lubm_heuristics_time}. Since the QDegree and Low-High heuristics touch and communicate almost the same amount of data, their cumulative execution times are also the same.
Besides, note that the QDegree heuristic does not use any statistical information from the data and only relies on the structure of the hot query pattern. Therefore, a redistributed pattern would not benefit other future queries with a slightly different structure. We repeated the experiment on WatDiv-1B and all heuristics resulted in almost the same communication cost, wall time, and touched data. This time, QDegree resulted in the least replication because its exploits best the initial subject-based hash partitioning.

\begin{table}[h]
\caption{Load Balancing in {\logo}}
\resizebox{\columnwidth}{!}{
\begin{tabular}{c|cccc|c}
\hline
\multirow{2}{*}{Dataset} & \multicolumn{4}{c|}{Percentage of triples} & \multicolumn{1}{c}{\multirow{2}{*}{\begin{tabular}[c]{@{}c@{}}Replication \\ Ratio\end{tabular}}} \\
                         & Max     & Min     & Average    & StDev ($\sigma$)    & \multicolumn{1}{c}{}                                                                              \\ \hline
LUBM-10240               &1.43\%   &1.35\%   & 1.39\%     & 0.02      &  0.73                                                                                                 \\
WatDiv-1B                &1.58\%   &1.20\%   & 1.33\%     & 0.07      &  0.36                                                                                                 \\ \hline
\end{tabular}
}
\label{tab:stat_lb}
\end{table}

\begin{figure}
\centering
  \subfigure[LUBM-10240]{
  	\label{fig:lubm_balance}
  	\includegraphics[width=0.64\columnwidth]{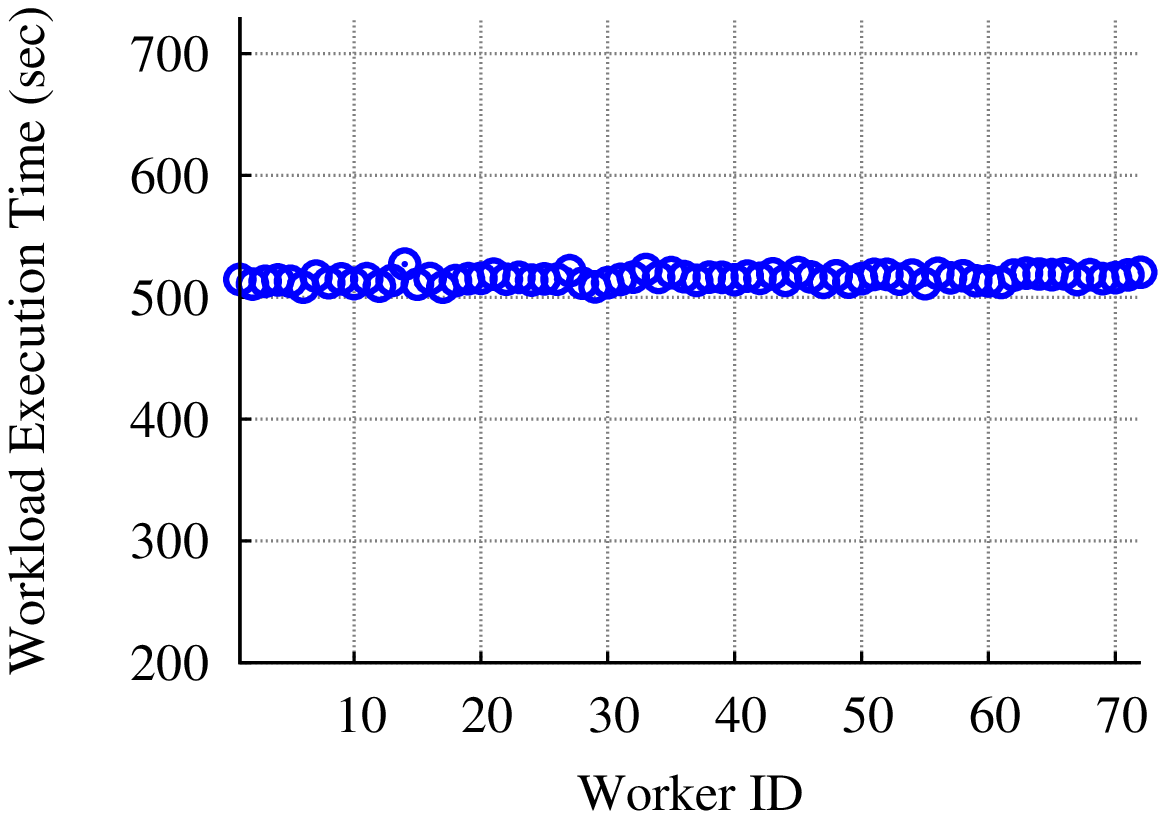}
  }
  \subfigure[WatDiv-1B]{
  	\label{fig:watdiv_balance}
  	\includegraphics[width=0.64\columnwidth]{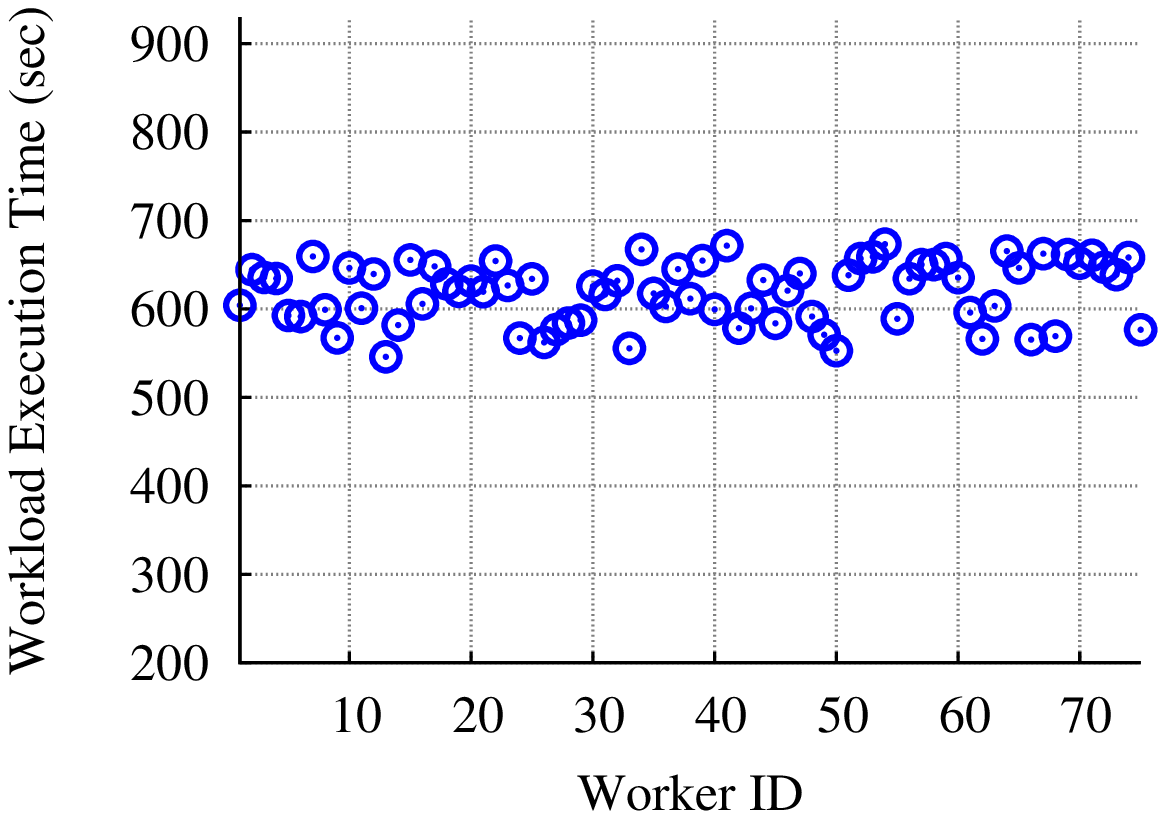}
  }
  \caption{Workload balance.}
\end{figure}

\begin{figure*}\centering
  \subfigure[Data scalability (simple)]{
  	\label{fig:scalability1}
  	\includegraphics[width=0.64\columnwidth]{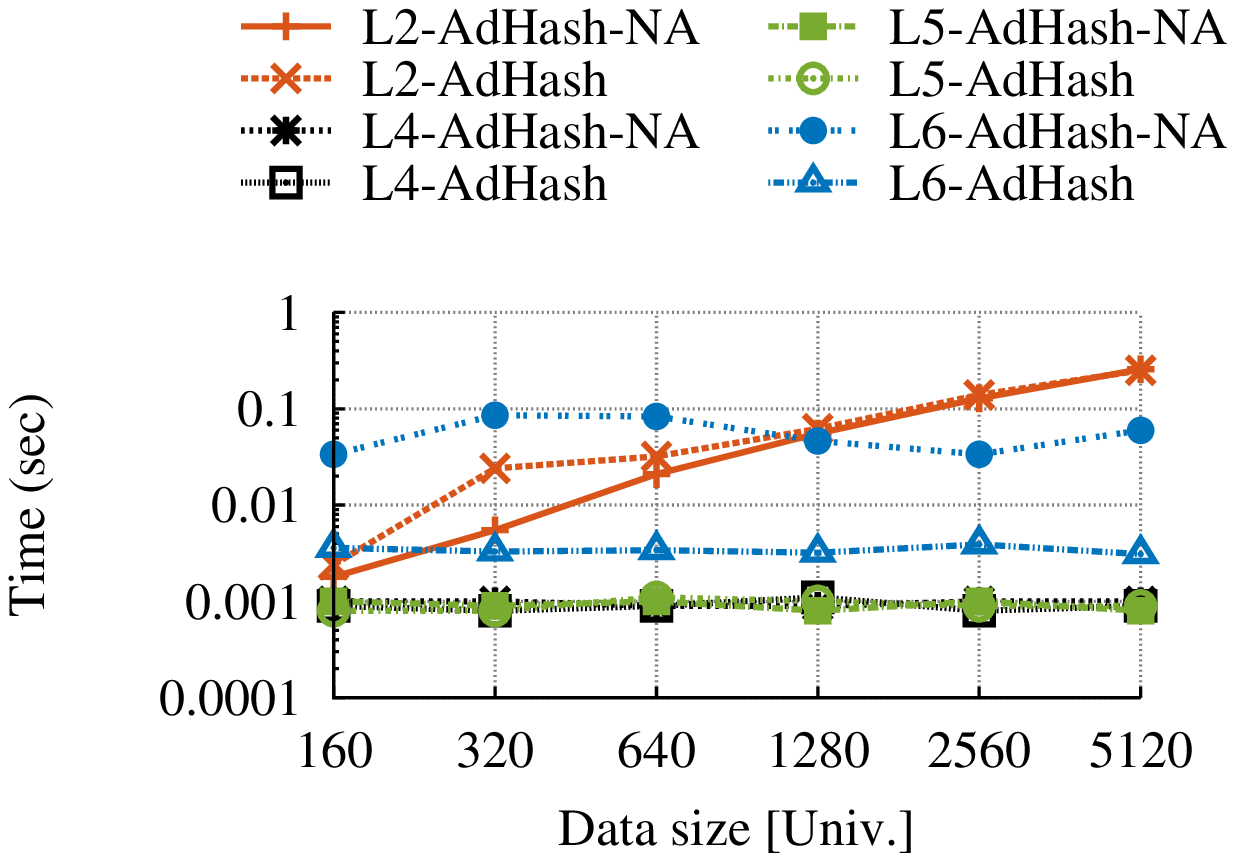}
  }\hfill
  \subfigure[Data scalability (complex)]{
  	\label{fig:scalability2}
  	\includegraphics[width=0.64\columnwidth]{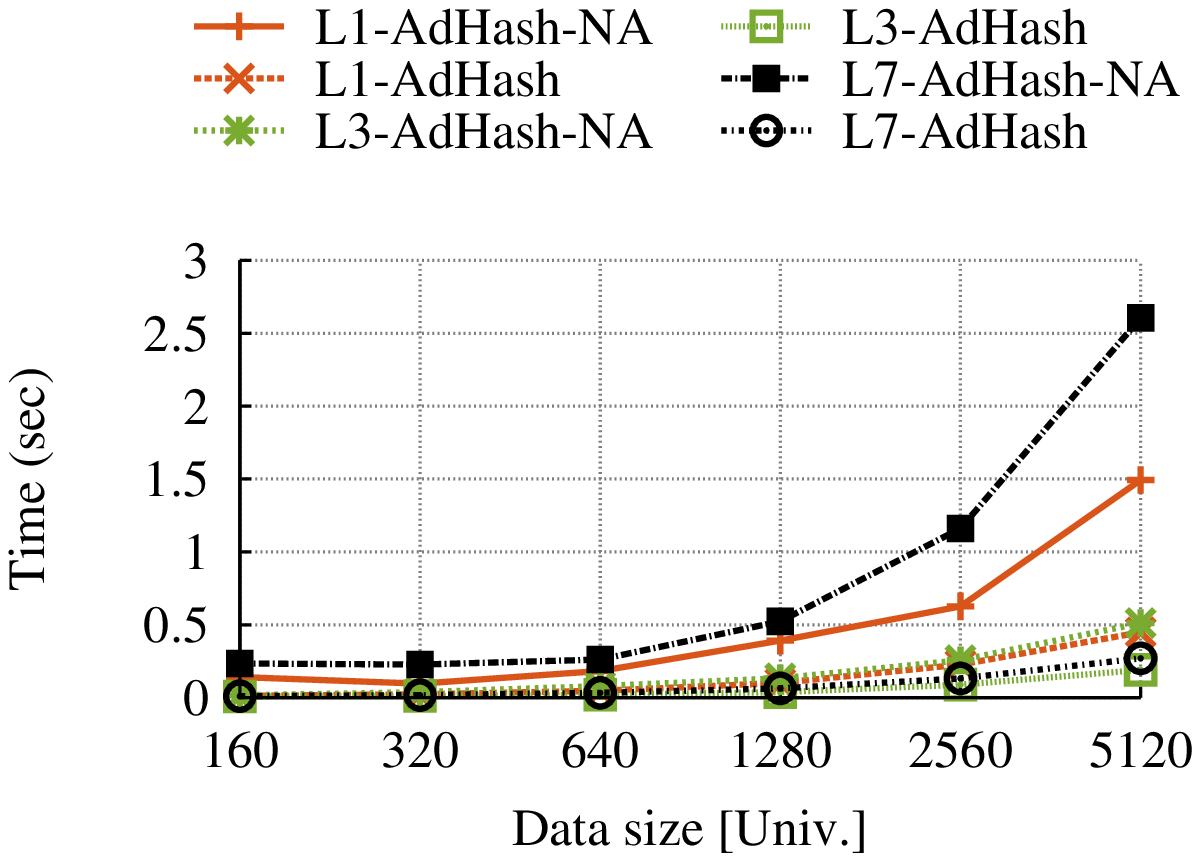}
  }\hfill
 \subfigure[Strong Scalability]{
  	\label{fig:scalability3}
  	\includegraphics[width=0.64\columnwidth]{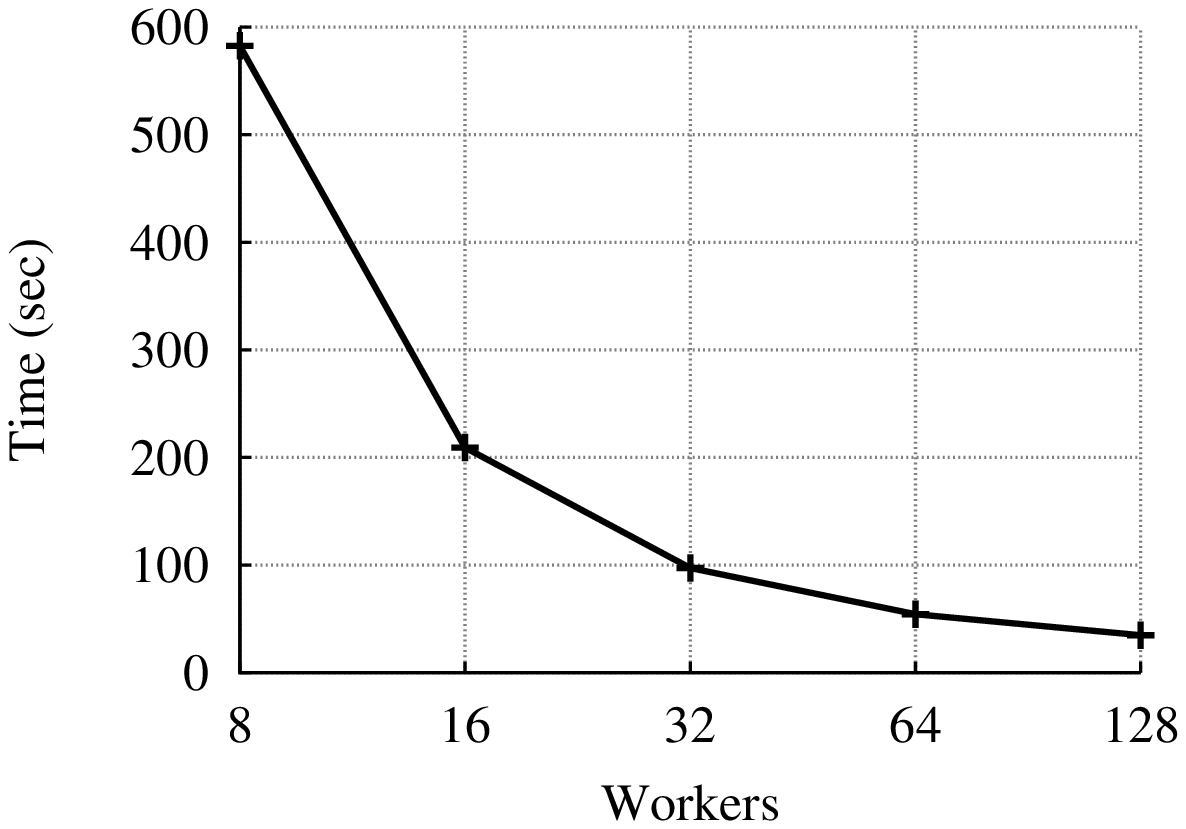}
  }
  \caption{{\logo} scalability using LUBM dataset. }
\end{figure*}

\subsubsection{Replication and Load Balancing}
\label{subsec:load_balancing}
In this experiment, we evaluate the load balancing of {\logo} from two different perspectives: \myNum{i} {\em data balancing}, in which we consider how balanced is the initial partitioning as well as the replication that results from the IRD process; {ii} {\em work balancing}, in which we consider how the evaluation cost is balanced among all workers in the system, during the execution of the workload. In Table \ref{tab:stat_lb}, we report some statistics that characterize the data load balance in {\logo}. Particularly, we report the average and standard deviation ($\sigma$) of the percentage of triples stored at each worker. As shown in the table, {\logo} achieves a very good data balance for both workloads because of the initial subject-based hash partitioning as well as the hashing used during the IRD process. As a result of the data balance, work is also well balanced among workers; i.e., the amount of work contributed by each worker in the system is almost the same as shown in Figures \ref{fig:lubm_balance} and \ref{fig:watdiv_balance} for the LUBM-10240 and WatDiv-1B, respectively.

\subsection{Scalability}
\label{subsec:scalable}

\stitle{Data Scalability}
We use LUBM benchmark data generator to generate six datasets of different sizes: LUBM-160, LUBM-320, LUBM-640, LUBM-1280, LUBM-2560 and LUBM-5120. 
We keep the number of workers fixed to 72 (6 workers per machine). Figures \ref{fig:scalability1} and \ref{fig:scalability2} shows the data scalability of {\logo} and {\logo}-NA for simple and complex queries respectively.
L4, L5, L6 are simple queries that are very selective and touch the same amount of data regardless of the data size. This describes the steady performance of both {\logo} and {\logo}-NA for these queries. Because L2 is not selective and returns massive final results, it is inevitable for its scalability to degrade as data size increases. 
Figure \ref{fig:scalability2} shows the scalability of {\logo} for complex queries. Queries L1 and L7 generate large number of intermediate results causing high communication cost,  
which explains their poor scalability of {\logo}-NA. Nevertheless, as {\logo} adapts to the workload, many queries are evaluated in parallel mode much faster.

\stitle{Strong Scalability}
Using LUBM-10240, we fixed the workload and increased the number of workers. Due to the adaptivity of {\logo}, communication is minimized leading to nearly optimal scalability. Figure \ref{fig:scalability3} shows the scalability of parallel queries as we increase the number of workers.

%% file: sec_conclusion.tex
\vspace{-25pt}
\section{Conclusion}
\label{sec:conclusion}

In this paper, we presented {\logo}, an adaptive distributed RDF engine. 
Using lightweight partitioning that hashes triples on the subjects, {\logo} 
exploits query structures and the hash-based data locality in order to
minimize the communication cost during query evaluation. 
Furthermore, {\logo} monitors the query workload and incrementally redistributes parts of the data that are frequently accessed by hot patterns. 
By maintaining and indexing these patterns, many future queries are evaluated without communication. 
The adaptivity feature of {\logo} complements its excellent performance on queries that can benefit from its hash-based data locality; i.e., frequent query patterns that are not favored by the partitioning (e.g., like star joins on an object) can be processed in parallel due to the {\logo}'s adaptivity.  

Our experimental results verify that {\logo} achieves better partitioning and replicates less data than its competitors. More importantly, {\logo} scales to very large RDF graphs and consistently provides superior performance by adapting to dynamically changing workloads. Currently, we are investigating the possibility of utilizing {\logo} for general (i.e., non-RDF) graphs, and operators such as graph traversals, or reachability queries.





%% file: sec_appendix.tex
\appendix

\section{LUBM Benchmark Queries}
\label{app:lubm_queries}
\footnotesize
{
PREFIX rdf: $<$http://www.w3.org/1999/02/22-rdf-syntax-ns\#$>$
\\
PREFIX ub: $<$http://www.lehigh.edu/~zhp2/2004/0401/univ-bench.owl\#$>$
\\
PREFIX rdfs: $<$http://www.w3.org/2000/01/rdf-schema\#$>$
\\
PREFIX y: $<$http://yago-knowledge.org/resource/$>$
}\\
\textbf{Q1:} SELECT ?X WHERE\{ 
    ?X rdf:type ub:GraduateStudent .
    \seqsplit{?X ub:takesCourse $<$http://www.Department0.University0.edu/GraduateCourse0$>$ .}
\}\\
\textbf{Q2:} SELECT ?X ?Y ?Z WHERE\{
    ?X rdf:type \seqsplit{ub:GraduateStudent} .
    ?Y rdf:type ub:University .
    ?Z rdf:type ub:Department .
    ?X ub:memberOf ?Z .
    ?Z ub:subOrganizationOf ?Y .
    ?X \seqsplit{ub:undergraduateDegreeFrom} ?Y .
\}\\
\textbf{Q3:} SELECT ?X	WHERE\{
    ?X rdf:type ub:Publication .
    ?X ub:publicationAuthor $<$\seqsplit{http://www.Department0.University0.edu/AssistantProfessor0}$>$ .
\}\\
\textbf{Q4:} SELECT ?X, ?Y1, ?Y2, ?Y3 WHERE{
    ?X rdf:type \seqsplit{ub:AssociateProfessor} .
    ?X ub:worksFor $<$\seqsplit{http://www.Department0.University0.edu}$>$ .
    ?X ub:name ?Y1 .
    ?X ub:emailAddress ?Y2 .
    ?X ub:telephone ?Y3 .
\}\\
\textbf{Q5:} SELECT ?X WHERE\{
    ?X rdf:type \seqsplit{ub:UndergraduateStudent} .
    ?X ub:memberOf $<$\seqsplit{http://www.Department0.University0.edu}$>$ .
\}\\
\textbf{Q6:} SELECT ?X WHERE\{
        ?X rdf:type \seqsplit{ub:UndergraduateStudent} .
\}\\
\textbf{Q7:} SELECT ?X, ?Y WHERE\{
    ?X rdf:type \seqsplit{ub:UndergraduateStudent} .
    ?Y rdf:type ub:Course .
    ?X ub:takesCourse ?Y .
    $<$\seqsplit{http://www.Department0.University0.edu/AssociateProfessor0}$>$ ub:teacherOf ?Y .
\}\\
\textbf{Q8:} SELECT ?X, ?Y, ?Z WHERE\{
    ?X rdf:type \seqsplit{ub:UndergraduateStudent} .
    ?Y rdf:type ub:Department .
    ?X ub:memberOf ?Y .
    ?Y ub:subOrganizationOf $<$\seqsplit{http://www.University0.edu}$>$ .
    ?X ub:emailAddress ?Z .
\}\\
\textbf{Q9:} SELECT ?X, ?Y, ?Z WHERE\{
    ?X rdf:type \seqsplit{ub:GraduateStudent} .
    ?Y rdf:type \seqsplit{ub:AssociateProfessor} .
    ?Z rdf:type ub:GraduateCourse .
    ?X ub:advisor ?Y .
    ?Y ub:teacherOf ?Z .
    ?X ub:takesCourse ?Z .
\}\\
\textbf{Q10:} SELECT ?X	WHERE\{
    ?X rdf:type \seqsplit{ub:TeachingAssistant} .
    ?X ub:takesCourse $<$\seqsplit{http://www.Department0.University0.edu/GraduateCourse0}$>$ .
\}\\
\textbf{Q11:} SELECT ?X WHERE\{
    ?X rdf:type ub:ResearchGroup .
    ?X ub:subOrganizationOf ?Z .
    ?Z ub:subOrganizationOf $<$\seqsplit{http://www.University0.edu}$>$ .
\}\\
\textbf{Q12:} SELECT ?X, ?Y WHERE\{
    ?Y rdf:type ub:Department .
    ?X ub:headOf ?Y.
    ?Y ub:subOrganizationOf $<$\seqsplit{http://www.University0.edu}$>$ .
\}\\
\textbf{Q13:} SELECT ?X WHERE\{
    ?X rdf:type ub:GraduateStudent .
    ?X ub:undergraduateDegreeFrom $<$\seqsplit{http://www.University0.edu}$>$ .
\}\\
\textbf{Q14:} SELECT ?X WHERE\{
        ?X rdf:type ub:GraduateStudent .
\}

\vspace{-20pt}
\section{LUBM Workload}
\label{app:lubm_wload}
We generated a workload of 20,000 queries from LUBM benchmark queries shown in \ref{app:lubm_queries}.  
For queries that do not have constants (Q2 and Q9), we generate different query patterns by removing some triples and mutating the node types. 
For example, in Q2, we generated 18 different patterns by alternating student type between UndergraduateStudent and GraduateStudent (see Table \ref{tab:lubmworkload}). 
Similarly, other query patterns are generated by removing different combinations of the query triple patterns. 
We did not generate variations of Q6 and Q14 as they have only one triple pattern (\emph{rdf:type}) with a single constant. For the rest of the queries, we generated 
1000 different patterns from each query by varying the values of the query constants. For example, in Q1, we generate different query patterns by varying the values of both student type (UndergraduateStudent or GraduateStudent) and graduate courses. 
\begin{table}[h]\scriptsize
\centering
\caption{LUBM Workload}
\begin{tabular}{l l l}
 \hline
     & \multicolumn{1}{c}{\textbf{Patterns}} & \multicolumn{1}{c}{\textbf{Changes}} \\ \hline
    Q1 	&1000	&Constants	   \\ 
    Q2 	&18		&Structure/Constants	   \\ 
    Q3 	&1000		&Constants	   \\ 
    Q4 	&1000		&Constants	   \\ 
    Q5 	&1000		&Constants	   \\ 
    Q6 	&1		&No Changes	   \\ 
    Q7 	&1000		&Constants	   \\  
    Q8 	&1000		&Constants	   \\ 
    Q9 	&30		&Structure/Constants	   \\ 
    Q10 	&1000		&Constants	   \\ 
    Q11 	&1000		&Constants	   \\ 
    Q12 	&1000		&Constants	   \\ 
    Q13 	&1000		&Constants	   \\ 
    Q14 	&1		&No Changes	   \\ \hline
    \end{tabular}
\label{tab:lubmworkload}
\end{table}

\vspace{-20pt}
\section{YAGO2 Queries}
\label{app:yago_queries}
\textbf{Y1:} SELECT ?GivenName ?FamilyName WHERE\{
    ?p \seqsplit{y:hasGivenName} ?GivenName . 
    ?p y:hasFamilyName ?FamilyName . 
    ?p \seqsplit{y:wasBornIn} ?city . 
    ?p y:hasAcademicAdvisor ?a .
    ?a y:wasBornIn ?city .
\}\\
\textbf{Y2:} SELECT ?GivenName ?FamilyName WHERE\{
    ?p \seqsplit{y:hasGivenName} ?GivenName . 
    ?p y:hasFamilyName ?FamilyName . 
    ?p \seqsplit{y:wasBornIn} ?city . 
    ?p y:hasAcademicAdvisor ?a .
    ?a y:wasBornIn ?city .
    ?p y:isMarriedTo ?p2 .
    ?p2 y:wasBornIn ?city .
\}\\
\textbf{Y3:} SELECT ?name1 ?name2 WHERE\{
    ?a1 \seqsplit{y:hasPreferredName} ?name1 . 
    ?a2 y:hasPreferredName ?name2 .
    ?a1 y:actedIn ?movie .
    ?a2 y:actedIn ?movie .
\}\\
\textbf{Y4:} SELECT ?name1 ?name2 WHERE\{
    ?p1 \seqsplit{y:hasPreferredName} ?name1 .
    ?p2 y:hasPreferredName ?name2 .
    ?p1 y:isMarriedTo ?p2 .
    ?p1 y:wasBornIn ?city .
    ?p2 y:wasBornIn ?city .
\}\\
%

\vspace{-20pt}
\section{Bio2RDF}
\label{app:bio_queries}
\textbf{B1:} SELECT ?o WHERE\{
    $<$\seqsplit{http://bio2rdf.org/pubmed\_resource:1374967\_INVESTIGATOR\_1}$>$ $<$\seqsplit{http://bio2rdf.org/pubmed\_vocabulary:last\_name}$>$ ?o .
    $<$\seqsplit{http://bio2rdf.org/pubmed\_resource:1374967\_AUTHOR\_1}$>$ $<$\seqsplit{http://bio2rdf.org/pubmed\_vocabulary:last\_name}$>$ ?o .
\}\\
\textbf{B2:} SELECT ?articleToMesh WHERE\{
$<$\seqsplit{http://bio2rdf.org/pubmed:126183}$>$ $<$\seqsplit{http://bio2rdf.org/pubmed\_vocabulary:mesh\_heading}$>$ ?articleToMesh .
?articleToMesh $<$\seqsplit{http://bio2rdf.org/pubmed\_vocabulary:mesh\_descriptor\_name}$>$ ?mesh .
\}\\
\textbf{B3:} SELECT ?phenotype WHERE\{
?phenotype rdf:type $<$\seqsplit{http://bio2rdf.org/omim\_vocabulary:Phenotype}$>$ .
?phenotype rdfs:label ?label .
?gene $<$\seqsplit{http://bio2rdf.org/omim\_vocabulary:phenotype}$>$ ?phenotype .
\}\\
\textbf{B4:} SELECT ?pharmgkbid WHERE\{
?pharmgkbid $<$\seqsplit{http://bio2rdf.org/pharmgkb\_vocabulary:xref}$>$ $<$\seqsplit{http://bio2rdf.org/drugbank:DB00126}$>$ .
?pharmgkbid $<$\seqsplit{http://bio2rdf.org/pharmgkb\_vocabulary:xref}$>$ ?pccid .
?DDIassociation $<$\seqsplit{http://bio2rdf.org/pharmgkb\_vocabulary:chemical}$>$ ?pccid .
?DDIassociation $<$\seqsplit{http://bio2rdf.org/pharmgkb\_vocabulary:event}$>$ ?DDIevent .
?DDIassociation $<$\seqsplit{http://bio2rdf.org/pharmgkb\_vocabulary:chemical}$>$ ?drug2 .
?DDIassociation $<$\seqsplit{http://bio2rdf.org/pharmgkb\_vocabulary:p-value}$>$ ?pvalue .
\}\\
\textbf{B5:} SELECT ?interaction WHERE\{
?interaction $<$\seqsplit{http://bio2rdf.org/irefindex\_vocabulary:interactor\_a}$>$ $<$\seqsplit{http://bio2rdf.org/uniprot:O17680}$>$ .
\}

%% file: paper.bbl
\begin{thebibliography}{10}
\providecommand{\url}[1]{{#1}}
\providecommand{\urlprefix}{URL }
\expandafter\ifx\csname urlstyle\endcsname\relax
  \providecommand{\doi}[1]{DOI~\discretionary{}{}{}#1}\else
  \providecommand{\doi}{DOI~\discretionary{}{}{}\begingroup
  \urlstyle{rm}\Url}\fi

\bibitem{wlm}
Alu\c{c}, G., \"Ozsu, M.T., Daudjee, K.: {Workload Matters: Why {RDF} Databases
  Need a New Design}.
\newblock PVLDB \textbf{7}(10) (2014)

\bibitem{atre}
Atre, M., Chaoji, V., Zaki, M.J., Hendler, J.A.: {Matrix "Bit" loaded: a
  scalable lightweight join query processor for RDF data}.
\newblock In: WWW (2010)

\bibitem{Blanas}
Blanas, S., Li, Y., Patel, J.M.: {Design and evaluation of main memory hash
  join algorithms for multi-core CPUs}.
\newblock In: SIGMOD (2011)

\bibitem{outliars}
Bol'shev, L., Ubaidullaeva, M.: {Chauvenet's Test in the Classical Theory of
  Errors}.
\newblock Theory of Probability \& Its Applications \textbf{19}(4), 683--692
  (1975)

\bibitem{moore}
Boyer, R.S., {Strother Moore}, J.: {MJRTY: A Fast Majority Vote Algorithm}.
\newblock In: R.S. Boyer (ed.) Automated Reasoning: Essays in Honor of Woody
  Bledsoe, pp. 105--118. Kluwer, London (1991)

\bibitem{Chong}
Chong, Z., Chen, H., Zhang, Z., Shu, H., Qi, G., Zhou, A.: {RDF pattern
  matching using sortable views}.
\newblock In: CIKM (2012)

\bibitem{schism10}
Curino, C., Jones, E., Zhang, Y., Madden, S.: {Schism: a workload-driven
  approach to database replication and partitioning}.
\newblock PVLDB \textbf{3}(1-2) (2010)

\bibitem{mapred}
Dean, J., Ghemawat, S.: Mapreduce: Simplified data processing on large
  clusters.
\newblock In: OSDI (2004)

\bibitem{hadooppp}
Dittrich, J., Quian{\'e}-Ruiz, J.A., Jindal, A., Kargin, Y., Setty, V., Schad,
  J.: {Hadoop++: Making a Yellow Elephant Run Like a Cheetah (Without It Even
  Noticing)}.
\newblock PVLDB \textbf{3}(1-2) (2010)

\bibitem{Dritsou}
Dritsou, V., Constantopoulos, P., Deligiannakis, A., Kotidis, Y.: {Optimizing
  query shortcuts in RDF databases}.
\newblock In: ESWC (2011)

\bibitem{recovery1}
Elnozahy, E.N.M., Alvisi, L., Wang, Y.M., Johnson, D.B.: {A Survey of
  Rollback-recovery Protocols in Message-passing Systems}.
\newblock ACM Comput. Surv. \textbf{34}(3), 375--408 (2002)

\bibitem{partout}
Galarraga, L., Hose, K., Schenkel, R.: {Partout: {A} Distributed Engine for
  Efficient {RDF} Processing}.
\newblock CoRR \textbf{abs/1212.5636} (2012)

\bibitem{empirical}
Gallego, M.A., Fern{\'a}ndez, J.D., Mart{\'\i}nez-Prieto, M.A., de~la Fuente,
  P.: {An empirical study of real-world SPARQL queries}.
\newblock In: USEWOD (2011)

\bibitem{Goasdoue}
Goasdou{\'e}, F., Karanasos, K., Leblay, J., Manolescu, I.: {View selection in
  Semantic Web databases}.
\newblock PVLDB \textbf{5}(2) (2011)

\bibitem{triad}
Gurajada, S., Seufert, S., Miliaraki, I., Theobald, M.: {TriAD: A Distributed
  Shared-nothing RDF Engine Based on Asynchronous Message Passing}.
\newblock In: SIGMOD (2014)

\bibitem{Harth07yars2}
Harth, A., Umbrich, J., Hogan, A., Decker, S.: {YARS2: A Federated Repository
  for Querying Graph Structured Data from the Web}.
\newblock In: ISWC/ASWC, vol. 4825 (2007)

\bibitem{warp}
Hose, K., Schenkel, R.: {WARP: Workload-aware replication and partitioning for
  RDF}.
\newblock In: ICDEW (2013)

\bibitem{Huang11:scalable}
Huang, J., Abadi, D., Ren, K.: {Scalable SPARQL Querying of Large RDF Graphs}.
\newblock PVLDB \textbf{4}(11) (2011)

\bibitem{hadooprdf}
Husain, M., McGlothlin, J., Masud, M., Khan, L., Thuraisingham, B.:
  {Heuristics-Based Query Processing for Large RDF Graphs Using Cloud
  Computing}.
\newblock TKDE \textbf{23}(9) (2011)

\bibitem{IdreosKM07}
Idreos, S., Kersten, M.L., Manegold, S.: {Database Cracking}.
\newblock In: CIDR (2007)

\bibitem{metis}
Karypis, G., Kumar, V.: {A Fast and High Quality Multilevel Scheme for
  Partitioning Irregular Graphs}.
\newblock SIAM J. Sci. Comput. \textbf{20}(1), 359--392 (1998)

\bibitem{shape}
Lee, K., Liu, L.: {Scaling Queries over Big RDF Graphs with Semantic Hash
  Partitioning}.
\newblock PVLDB \textbf{6}(14) (2013)

\bibitem{pregel10}
Malewicz, G., Austern, M., Bik, A., Dehnert, J., Horn, I., Leiser, N.,
  Czajkowski, G.: {Pregel: a System for Large-scale Graph Processing}.
\newblock In: SIGMOD (2010)

\bibitem{Neumann2010}
Neumann, T., Weikum, G.: The rdf-3x engine for scalable management of rdf data.
\newblock VLDB J. \textbf{19}(1), 91--113 (2010)

\bibitem{h2rdf2}
Papailiou, N., Konstantinou, I., Tsoumakos, D., Karras, P., Koziris, N.:
  H2rdf+: High-performance distributed joins over large-scale rdf graphs.
\newblock In: IEEE Big Data (2013)

\bibitem{rya}
{Punnoose, Roshan and Crainiceanu, Adina and Rapp, David}: {Rya: A Scalable RDF
  Triple Store for the Clouds}.
\newblock In: Cloud-I (2012)

\bibitem{sampled}
Rietveld, L., Hoekstra, R., Schlobach, S., Gu{\'e}ret, C.: {Structural
  Properties as Proxy for Semantic Relevance in RDF Graph Sampling}.
\newblock In: ISWC (2014)

\bibitem{Rohloff10:shard}
Rohloff, K., Schantz, R.E.: {High-performance, massively scalable distributed
  systems using the MapReduce software framework: the SHARD triple-store}.
\newblock In: PSI EtA (2010)

\bibitem{recovery2}
Shen, Y., Chen, G., Jagadish, H.V., Lu, W., Ooi, B.C., Tudor, B.M.: {Fast
  Failure Recovery in Distributed Graph Processing Systems}.
\newblock PVLDB \textbf{8}(4) (2014)

\bibitem{stonebraker07}
Stonebraker, M., Madden, S., Abadi, D., Harizopoulos, S., Hachem, N., Helland,
  P.: {The end of an Architectural Era: (It's Time for a Complete Rewrite)}.
\newblock In: PVLDB (2007)

\bibitem{billion}
Wang, L., Xiao, Y., Shao, B., Wang, H.: {How to partition a billion-node
  graph}.
\newblock In: ICDE (2014)

\bibitem{Weiss2008}
Weiss, C., Karras, P., Bernstein, A.: {Hexastore: sextuple indexing for
  semantic web data management}.
\newblock PVLDB \textbf{1}(1) (2008)

\bibitem{path}
{Wu, Buwen and Zhou, Yongluan and Yuan, Pingpeng and Liu, Ling and Jin, Hai}:
  {Scalable SPARQL Querying using Path Partitioning}.
\newblock In: ICDE (2015)

\bibitem{yang12}
Yang, S., Yan, X., Zong, B., Khan, A.: {Towards effective partition management
  for large graphs}.
\newblock In: SIGMOD (2012)

\bibitem{triplebit}
Yuan, P., Liu, P., Wu, B., Jin, H., Zhang, W., Liu, L.: {TripleBit: A Fast and
  Compact System for Large Scale RDF Data}.
\newblock PVLDB \textbf{6}(7), 517--528 (2013)

\bibitem{spark}
Zaharia, M., Chowdhury, M., Franklin, M.J., Shenker, S., Stoica, I.: {Spark:
  Cluster Computing with Working Sets}.
\newblock In: USENIX (2010)

\bibitem{trinity.rdf}
Zeng, K., Yang, J., Wang, H., Shao, B., Wang, Z.: {A distributed graph engine
  for web scale RDF data}.
\newblock PVLDB \textbf{6}(4) (2013)

\bibitem{eagre}
Zhang, X., Chen, L., Tong, Y., Wang, M.: {EAGRE: Towards scalable I/O efficient
  SPARQL query evaluation on the cloud}.
\newblock In: ICDE (2013)

\bibitem{gstore}
Zou, L., \"{O}zsu, M.T., Chen, L., Shen, X., Huang, R., Zhao, D.: {gStore: A
  Graph-based SPARQL Query Engine}.
\newblock VLDB J. \textbf{23}(4), 565--590 (2014)

\end{thebibliography}
